\documentclass{article}

\usepackage{cite}
\usepackage{PRIMEarxiv}

\usepackage[utf8]{inputenc} % allow utf-8 input
\usepackage[T1]{fontenc}    % use 8-bit T1 fonts
\usepackage{hyperref}       % hyperlinks
\usepackage{url}            % simple URL typesetting
\usepackage{booktabs}       % professional-quality tables
\usepackage{amsfonts, amsmath, amsthm, amssymb} % For math fonts, symbols and environments
\usepackage{nicefrac}       % compact symbols for 1/2, etc.
\usepackage{microtype}      % microtypography
\usepackage{lipsum}
\usepackage{fancyhdr}       % header
\usepackage{graphicx}       % graphics
\graphicspath{{media/}}     % organize your images and other figures under media/ folder
\usepackage[font=small,labelfont=bf]{caption} % make figure label bold and make the caption content small
\usepackage{subcaption}
\usepackage[ruled]{algorithm2e}

\title{Scalable Agent-Based Modeling for Complex Financial Market Simulations 
}

\author{
  Aaron Wheeler \\
  R.F Smith School of Chemical and Biomolecular Engineering \\
  Cornell University \\
  Ithaca, NY \\
  \texttt{aw843@cornell.edu} \\
   \And
  Jeffrey D. Varner \\
  R.F Smith School of Chemical and Biomolecular Engineering \\
  Cornell University \\
  Ithaca, NY \\
  \texttt{jdv27@cornell.edu} \\
}

\begin{document}
\maketitle

\begin{abstract}
In this study, we developed a computational framework for simulating large-scale agent-based financial markets. Our platform supports trading multiple simultaneous assets and leverages distributed computing to scale the number and complexity of simulated agents. Heterogeneous agents make decisions in parallel, and their orders are processed through a realistic, continuous double auction matching engine. We present a baseline model implementation and show that it captures several known statistical properties of real financial markets (i.e., stylized facts). Further, we demonstrate these results without fitting models to historical financial data. Thus, this framework could be used for direct applications such as human-in-the-loop machine learning or to explore theoretically exciting questions about market microstructure's role in forming the statistical regularities of real markets.  To the best of our knowledge, this study is the first to implement multiple assets, parallel agent decision-making, a continuous double auction mechanism, and intelligent agent types in a scalable real-time environment. 
\end{abstract}

% keywords can be removed

\section{Introduction}
% \subsection{Motivation}
Modern society and the global economy heavily rely on financial markets, making it essential to develop tools that can help understand and navigate their complexity.  Beyond the analysis of historical data sets, which are costly and limited, we focus here on methods for generating and interacting with synthetic data---i.e., realistic market simulations. In addition to its low cost and accessibility, simulation also benefits from the ability to incorporate feedback effects and generate data for scenarios that have occurred only in rare circumstances---or perhaps scenarios that have yet to happen. High-fidelity market simulators can benefit market stakeholders, including retail investors, institutional investors, regulators, and others, by testing execution algorithms across various scenarios, developing more robust trading models, training models to detect financial crime, and observing policy changes' effects before implementing new regulations. Therefore, a realistic market simulation technology can promote more efficient, fair, and stable markets, benefiting the public. 

% \subsection{Market structure and stylized facts}
The modeling of financial markets is challenging due to various factors, including the large number of participants in the market, their adaptability and connectedness, and the fast pace of market events. Modelers often abstract market features from their models to deal with this complexity. Further, it is often difficult to reason which features are necessary or unnecessary. Market microstructure (i.e., the detailed description of trading mechanics \cite{OHara1997-wk}) are typically abstracted away in models. For example, in a real equities market, shares are bought and sold on electronic exchanges through digital records known as order books. Order books hold the queue of pending orders for a particular asset, and orders are fulfilled and removed from the book using matching algorithms (e.g., price-time priority). Thus, order books are a real-time record of supply and demand for a particular asset. On the demand side, buyers submit \texttt{bids} while sellers submit \texttt{asks} on the supply side of the book. The \texttt{bid-ask spread} is the difference between the best bid and the best ask price. Relatedly, \texttt{liquidity} measures how easy it is to carry out a transaction at a stable price; illiquid assets have sparse order books with a wide bid-ask spread. Crucially, market-maker firms provide liquidity by maintaining concurrent offers to conduct transactions between buyers and sellers. Without the inclusion of order books and market-makers, financial models fail to capture empirical regularities of markets and have limited use in practice \cite{Paulin2018, Gould2013, Bouchaud2008}.

Empirical regularities of financial markets---i.e., nontrivial statistical properties that are observed across a wide range of instruments, markets, and periods ---are known as \texttt{stylized facts} \cite{Cont2001, Chakraborti2011, Gould2013, Vyetrenko2020}. Research on stylized facts can be traced back to the 1960s when Mandelbrot studied the heavy-tailed distributions and long-term correlations of financial price return time series \cite{Mandelbrot1963, mandelbrot1971analysis}. Over time, other distinct stylized facts have been identified and categorized as univariate properties (related to a single asset) or multivariate properties (related to multiple assets). The most extensively researched phenomena are the univariate stylized facts that describe the characteristics seen in both price and volume series of a single asset in isolation \cite{Cont2001, Chakraborti2011, Gopikrishnan1999, Engle1982, Cont2005, Siven2009}. Univariate stylized facts also describe regularities found on short timescales such as order flow patterns and other microstructural phenomena \cite{Gould2013, Bouchaud2002}. Multivariate stylized facts involve phenomena regarding a basket of assets and the relationships between them \cite{Chakraborti2011, Stock1988, Plerou2002}. Stylized facts are interesting from a modeling perspective because they can be used to gauge the efficacy of models; ideally, one would want a model that can generate synthetic data that captures all the information and properties of empirical time series. Consequently, the collection of stylized facts provides a generalized measure for how a synthetic time series captures the actual underlying market dynamics, i.e., a test for simulation fidelity.

\subsection{Related work}
The modeling and analysis of financial time series is a problem studied across several decades and by many different disciplines \cite{Sornette2014, Jiang2019}. Typically, one of two methods is adopted: a top-down or bottom-up approach. Traditional financial models, such as those based on mathematical processes like geometric Brownian motion (GBM), adopt a top-down approach where macroscopic properties (e.g., asset prices) are modeled directly. In a top-down approach, the underlying dynamic market microstructure is described by a small number of lumped parameters \cite{Cox1979, Engle2001, Harvey1990, Stock2001}. Typically, empirical data is used to estimate these parameters. Then, simulation approaches like Monte Carlo methods can be applied for several purposes, including forecasting future outcomes and determining "fair prices" to assist investment decision-making \cite{Black1973}. However, top-down methods struggle to model market dynamics such as contagion effects, feedback loops, and structural changes. Further, top-down approaches are ill-suited for producing emergent phenomena (e.g., sudden price shocks and price discovery). 

Researchers have linked these highly nonlinear and interconnected market dynamics to properties of complex systems, which have been successfully modeled via bottom-up approaches \cite{axtell2022agent, Arthur2021, Bouchaud2021}. Agent-based models (ABMs) are a classical approach emphasizing the collective result of individual decisions and order flow to create macro-scale features from micro-scale interactions. ABMs have a long development history and have been utilized in modeling complex financial systems \cite{Chen2012}. ABMs have five critical components: software agents, interaction mechanism, simulation environment, calibration scheme, and validation procedure.

% Agent-based models (ABMs) implement a bottom-up approach that models the microscopic behaviors (e.g., individual decisions and order flows) and relies on the macro-scale features to emerge naturally as a byproduct of the micro-scale interactions. The modeling of complex financial systems as ABMs has a rich history \cite{Chen2012}, but it has yet to be established as either a rigorous engineering discipline or an effective tool in practice. However, there has been significant progress toward this aim in recent years, primarily due to the growth in computational power. As a modeling paradigm, ABMs comprise five components: software agents, an interaction mechanism, the simulation environment, a calibration scheme, and a validation procedure.

\begin{description}
   \item[Agents] Software agents are autonomous computational entities that act according to pre-defined rules and objectives. The agents can be heterogeneous, i.e., they can be defined with different properties and exhibit fundamentally different behaviors. Researchers have taken several different approaches to model the behavior of the agents. In the economics literature, agents have been modeled as rational investors who make decisions using strategies that maximize personal utility \cite{vonneumann1947, LeBaron1999}. On the other end of the spectrum, zero-intelligence approaches for agent behavior assume that orders are governed by stochastic processes \cite{FarmerZI2003, Ponta2011}. % mention population sizes of previous studies, tradeoff between simulation scale and agent intelligence?
   \item[Interaction] The interaction mechanism of an ABM dictates how agents interact with one another through both their rules (actions they are allowed to take) and the explicit structure to which they are linked. In the literature, approximations for the order-matching process are typically used. Examples include equations that balance all orders and set a clearing equilibrium price \cite{LeBaron1999, Raberto2001}, and machine learning models trained on empirical data \cite{Kuo2021, Ardon2021}. There are comparatively few examples of ABMs that use continuous double auction matching engines \cite{ABIDES2019, Lussange2022, Brandouy2013, Jericevich2021}, the actual mechanism employed by the New York Stock Exchange and Nasdaq. The reason for this absence lies in the inherent difficulty in simulating agents that make decisions with a range of available price options---and then aggregate to balance supply and demand forces (liquidity) sustainably \cite{Gould2013}.
   \item[Environment] The simulation environment defines the state of the simulation. This includes the observable variables, i.e., information available to some or all agents, and external factors, such as external shocks, etc. A more nuanced aspect of the simulation environment is the representation of time. Financial ABMs are typically represented in discrete time, where agents make decisions one after the other in turn-based games. % mention multple assets, fragmented exchange environment?
   \item[Calibration] To generate realistic behavior, ABMs typically undergo both a calibration and validation procedure. To calibrate the model, model parameters are configured by heuristic methods, initializing the model with empirical microscopic data, etc. Recently, researchers have proposed methods for using machine learning for calibrating ABMs \cite{Lamperti2017, Glielmo2023, Dahlke2020}. % discuss how some have used real data as an "oracle" etc.?
   \item[Validation] To validate the model, statistical tests can be run on the outputs and measured against what is known about the system, or the outputs can be measured against empirical macroscopic data. Regarding financial ABMs, validation is conducted by analyzing the model outputs and how they compare against known statistical properties of financial markets (i.e., stylized facts). % dicuss how most will use subset of stylized facts for validation rather than all?
\end{description}

This study presents a market simulation that uses an agent-based approach and supports a large population of agents, multiple assets, and computationally demanding agents. To achieve this, we distribute the various subsystems of our financial ABM platform across multiple machines. The simulation includes several agent types, most of which are based on zero-intelligence trading strategies. The platform uses a realistic continuous double auction mechanism with multiple order types and operates in real-time, making it useful for applications such as human-in-the-loop machine learning and demonstrating a model’s execution and performance in real-time. Moreover, it could be used to examine the impact of market microstructures on the formation of statistical regularities. The simulation was able to replicate several, but not all, of a set of typical stylized facts without using historical data for calibration. Although this baseline implementation produced statistically interesting time series data, there are several areas for improvement. However, we will leave extensions and optimization of the order processing, network communication, and other subsystems for future work.

\section{Materials and Methods}
\label{sec:headings}

\subsection{Construction of simulated market environment}
The ABM has a centralized processing engine and four unique agent types. Rather than fit our agent or environmental models to empirical data, model parameters were derived from first principles. We explicitly accounted for known components of financial markets---and made minimal assumptions elsewhere. For example, we used an explicit order book structure and matching engine mechanism. We also recreated nuanced features of natural market environments, such as multiple assets, parallel decision-making, asynchronous order flow, uncertain order execution, and potentially irrational agent behaviors. This reasoning is in line with the other zero-intelligence approaches in the literature \cite{FarmerZI2003, Ponta2011}, except this study is the only one to our knowledge that attempts this approach in an environment that includes all aforementioned market features. The simulation code was written in the Julia Programming Language for its high-performance and scientific computing capabilities \cite{Bezanson2017}. 

\begin{figure}[t]
    \centering
    \includegraphics[width=0.99\textwidth]{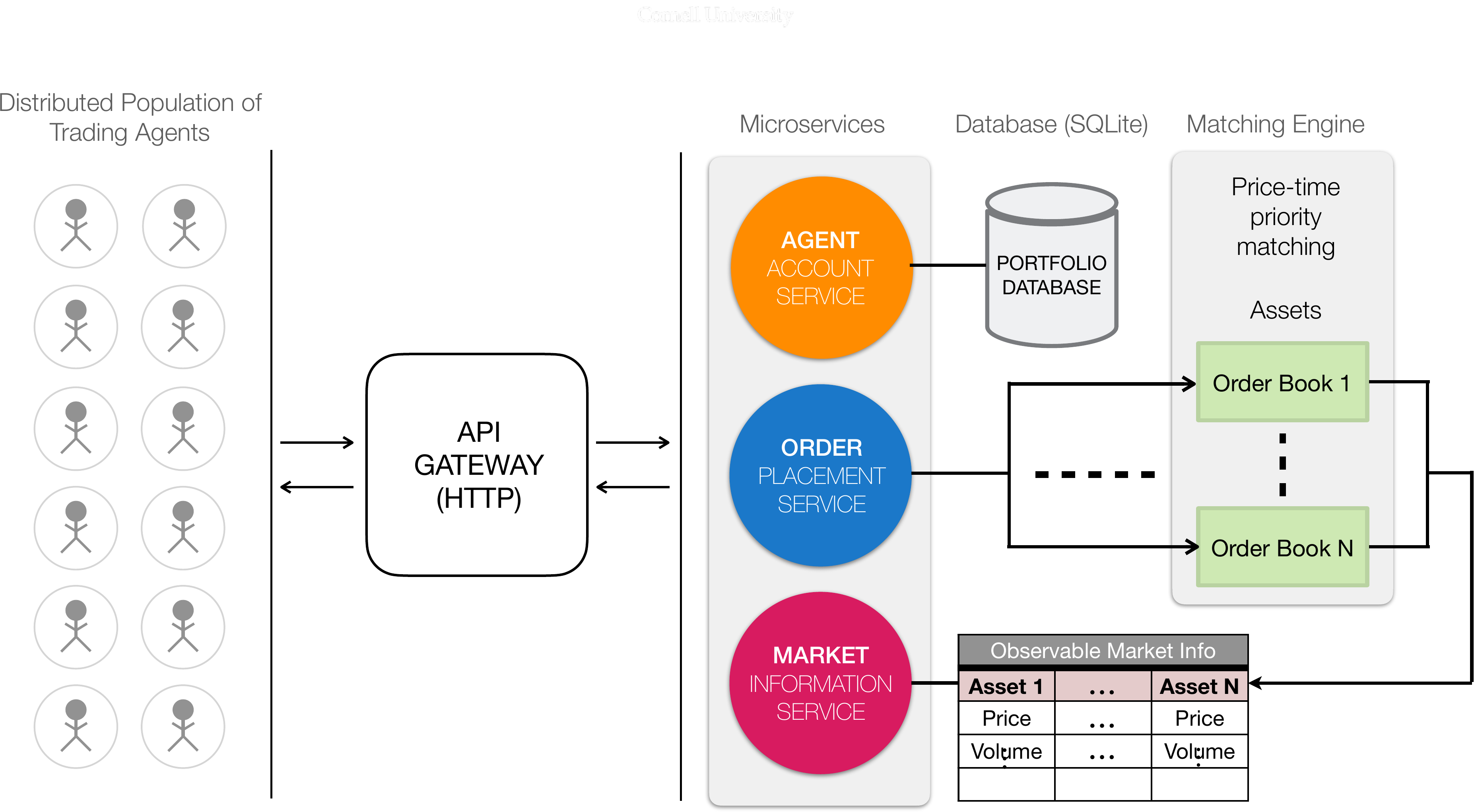}
    \caption{Schematic of the baseline financial agent-based market simulation architecture. We used a client-server architecture relying on the \texttt{HTTP} protocol for communication where clients (i.e., agents/market participants) communicated with a server (i.e., brokerage/exchange) to fulfill various services (e.g., query account information, submit orders, query market information, etc.). Agent decision-making and order processing were computationally decoupled, which allowed for a realistic setting where trading decisions were made in parallel and orders were processed serially.}
    \label{fig:sim_arch}
\end{figure}

The overall simulation consisted of several programs that intercommunicated to simulate market dynamics (Fig. \ref{fig:sim_arch}). All market participants (i.e., retail traders, fundamental investors, market makers, etc.) submitted orders in real time over a network via the \texttt{HTTP} protocol to the central system that captured this continuous data feed, facilitated trades and hosted the order books. The central system also hosted a market information server that can be queried on demand at any continuous time $t$ by all agents. % discuss how our initial design of this system operates more as a Brokerage than a stock exchange...

\paragraph{Observation space} Public market information (available to all agents) includes the best bid $p^{bid}_t$ and ask $p^{ask}_t$ prices, and the current cumulative trading volume $V_{total, t}$. From the bid-ask prices, agents can compute the mid-price defined as $p_{mid, t} = (p^{bid} + p^{ask}) / 2$. Further,  the price history of any asset was also available for all agents to query. In addition, order book information such as the depth of price levels, the number of unique orders at each price level, and the bid or ask side volumes were also available. However, this information was not used by all agents. Private agent-specific details such as available cash, active orders sitting in the order book, and the portfolio composition were restricted to the respective individual agents. Finally, identifying information was also withheld from other market participants, e.g., there is no way to determine who an order was submitted by. Thus, there was no way for an agent to learn a specific individual's trading pattern and try to predict what they'll buy/sell next.

\paragraph{Action space} The action space was defined by the discrete actions: (i) The price at which to place a buy order (provided by the use of a \emph{limit order}); (ii) The price at which to place a sell order (provided by the use of a \emph{limit order}); (iii) The amount of inventory to immediately trade or hedge (provided by the use of a buy or sell \emph{market order}); (iv) The amount of pending order to cancel (provided by the use of a \emph{cancel order}); (v) The option of doing nothing.

Specific agents were endowed with unique privileges; notably, market-making agents were allowed to possess negative inventories and negative cash, which was intentional and realistic due to margin trading and short selling, respectively. All other agent types cannot spend money they don't have and cannot trade shares they don't own. Finally, there was no influx of new cash or shares in the simulation, and the total initial amount of cash and shares was conserved.

\subsection{Agents}
We used four agent groups to generate trading activity in our simulated market environment: liquidity takers, liquidity providers, market-making agents, and intelligent agents. For simplicity, we avoided using some of the more specific agent types found in the literature, e.g., fundamental, technical, contrarian, etc. \cite{Chen2012}. Rather, we considered the liquidity takers and providers as an abstraction of these more specific agents types and used these groups as a stochastic process for modeling supply and demand through market, limit, and cancel orders. The exception to this is the market-making and intelligent agents, which were modeled to represent a specific type of actual market participant, market intermediaries (e.g., market makers, wholesalers, etc.). The pseudo-code for the four agent types in the simulation environment is provided in Appendix A.

\subsubsection{Liquidity takers}
The liquidity-taking agents were modified from previous work \cite{Ponta2011}, and they were the primary drivers of immediate price movement in the simulation---they removed the best pending orders from the book by submitting market orders exclusively. These agents were activated once the market was open according to a trading frequency parameter. Upon activation time $t$, the activated agent determined how much of their total wealth to devote to risky assets, i.e., they determined their risky wealth $\hat{\mathcal{W}}^{R}_{t}$:
\begin{eqnarray}
    \hat{\mathcal{W}}^{R}_{t} = \mathcal{X}_t(c_t + \sum_{k=1}^{K} h^k_t\cdot p^k_{mid, t}),
\end{eqnarray}
where $\mathcal{X}_t\sim U\left( 0,1 \right)$ denotes a random draw from a uniform distribution, $c_t$ denotes the agent's current cash, $h_t$ denotes the agent's current holdings (number of shares) of asset $k$, and $K$ denotes the total number of unique assets. Once the agent's risky wealth was determined, a Dirichlet distribution with a concentration parameter of 1.0 and $K$ categories was used to randomly determine the new desired portfolio weights $\hat{\omega}_{t+1}^k$. Using these new portfolio weights, the agent calculated the number of desired holdings of each asset $k$:
\begin{eqnarray}
    \hat{h}_{t+1}^k = \left\lfloor \hat{\omega}_{t+1}^k \cdot
    \frac{\hat{\mathcal{W}}^R_{t}}{p_{mid, t}^k} \right\rfloor
\end{eqnarray}
The corresponding order amount $\Delta^k_t$ was given by:
\begin{eqnarray}
    \Delta^k_t = \hat{h}_{t+1}^k - h_t^k
\end{eqnarray}
where $\Delta^k_t > 0$ resulted in a buy market order and $\Delta^k_t < 0$ resulted in a sell market order. Each activated agent repeated this process with their respective observations and properties until market close. These agents are highly scalable and can operate in parallel due to the simplicity of their trading logic and the ability to operate without the need to retain variables in local memory (all variables were instead queried from the brokerage server database at time $t$).

\subsubsection{Liquidity providers}
The liquidity-providing agents provided price movement resistance to the market---they added pending orders to the book by submitting limit orders exclusively. Identical to liquidity takers, these agents were activated according to a trading frequency parameter between market open and close times. Upon activation time $t$, the activated agent were assigned a limit price $p_t^{L_k}$ for each asset $k$ by:
\begin{eqnarray}
    p_t^{L_k} = p_{mid, t}^k (1+\mathcal{X}),
\end{eqnarray}
where $\mathcal{X}\sim \mathcal{N}\left(0,\sigma_{p_{mid, t}^k} \right)$ denotes a random draw from a normal distribution with the variance given by a volatility estimate $\sigma_{p_{mid, t}^k}$:
\begin{eqnarray}
    \sigma_{p_{mid, t}^k} = \sqrt{\frac{1}{T} \sum_{t=1}^{T} (r_t - \mu)^2}
\end{eqnarray}
where $r_t$ denotes the returns series, $r_t = ln(p_{t}) - ln(p_{t-1})$, and $\mu$ denotes the mean of the returns series. The agent made trading decisions based on the value of $p_t^{L_k}$. If $p_t^{L_k} > p_{mid, t}^k$ then the agent would issue a sell limit order for the asset $k$ in their portfolio, and if $p_t^{L_k} < p_{mid, t}^k$ then the agent would use their excess cash and issue a buy limit order to add the asset $k$ to their portfolio. Each activated agent repeated this exercise with their respective observations and properties until market close. Similar to liquidity takers, these agents are highly scalable and can operate in parallel due to the simplicity of their trading logic and the ability to operate without the need to retain variables in local memory.

\subsubsection{Market making agents}
The market-making agents facilitated trades by placing concurrent orders in the order book and were both providers and takers of liquidity in the simulation. As soon as the market was open, these agents would observe current market conditions and determine the price at which they would place concurrent bid and ask limit orders: 
%$p^{bid}_t = p_{mid, t} + S_{ref,t} (1 + \epsilon_{buy})$ and $p^{ask}_t = p_{mid, t} - S_{ref,t} (1 + \epsilon_{sell})$,
\begin{eqnarray}
    p^{bid}_t = p_{mid, t} + S_{\text{ref},t} (1 + \epsilon_{buy})\\
    p^{ask}_t = p_{mid, t} - S_{\text{ref},t} (1 + \epsilon_{sell})
\end{eqnarray}

where $\epsilon_{buy},\epsilon_{sell} \sim U\left(\epsilon_{min},\epsilon_{max} \right)$ denotes random draws from a uniform distribution with the range given by model parameters $\epsilon_{min},\epsilon_{max}$. In general, the price tweak factors $\epsilon$ can be understood as indicators of how passive or aggressive the order is; a trade with $\epsilon > 0$ is more passive since it is deeper into the book, and a trade with $\epsilon < 0$ is more aggressive since it is quoted at a more competitive price than the pre-existing best quote in the book. After the concurrent limit orders were routed to the central exchange server, the agent would determine how much of their current inventory to hedge. The agent determined the size of the hedge order by multiplying the size of their current inventory $z$ by a random fraction $\mathcal{X}_t\sim U\left( 0,1 \right)$. If $z > 0$, they would submit a sell market order; if $z < 0$ (i.e., they were in a short position), they would submit a buy market order. The agent would pause to allow the order to be filled, and then they would cancel any unfilled orders and update local variables. The market-making agents repeated this process according to a trading frequency parameter until market close time.

Compared to the liquidity provider and taker agents, the market-making agents are not as scalable due to the speed at which they operate. It is impractical to store their local data (i.e., cash and shares) because they would need to query it so often. Fortunately, the population size of this agent group is low relative to the liquidity provider and taker agents. % To simulate parallelism with this agent group, we have the ability to operate it on a separate machine.

\subsubsection{Intelligent agents}
The intelligent agents were included to mitigate excessively volatile market conditions, as previous work has suggested that strategic behavior is necessary to stabilize trading in order book-driven markets \cite{Chakraborti2011a}. Intelligent agents were modeled as adaptive market makers and learned to maximize profit while accounting for risk. They used two separate policies to accomplish this: a pricing policy and a hedging policy. The policies were inspired by Ganesh et al. (\cite{Ganesh2019}); each policy utilized estimates of the mean and variance of both the incoming net order flow $\nu_{\epsilon}$ and normalized spread profit-and-loss $s_{\epsilon}$ as a function of $\epsilon^{bid/ask}$. The incoming net order flow $\nu_{\epsilon}$ can be considered the amount of a placed order that is executed. The normalized spread profit-and-loss is defined as $s_{\epsilon} = (\nu S^{bid/ask}_{t}) / S_{ref,t}$ for $S^{bid/ask}_{t} = S_{ref,t} (1 + \epsilon^{bid/ask})$ and trade size $\nu$.

\paragraph{Empirical estimates}
The intelligent agent relied on hand-crafted optimization procedures to execute trading actions for optimal market-making in real-time. These procedures hinged on conditional expectation values $\mathbb{E}(\nu_{\epsilon})$ and $\mathbb{E}(s_{\epsilon})$. For practical reasons, the intelligent agent estimated these values using a linear estimator. Specifically, the estimates were assumed to follow a process given by:
\begin{align*}
    & y = A x + e
    \\
    & \hat{x} = (A^T A)^{-1} A^T y
\end{align*}
where $\hat{x}$ denotes the Least-squares estimator and $e$ denotes a Gaussian noise variable with $\mathbb{E}(e) = 0$. For the net order flow $\nu_{\epsilon}$, the observed label $y_{\nu} \in \mathbb{R}^m$ and the design matrix of given features $A_{\nu} \in \mathbb{R}^{m \times 4}$ was formulated as:
%\begin{align*} 
%    y = \nu_{\epsilon} \quad \quad A_{\nu} = \left[1.0,p_{mid,t},S_{ref,t},\epsilon\right]
%\end{align*}

\begin{align*}
    y_{\nu} = 
    \begin{bmatrix}
           \nu_{\epsilon_{t_0,buy}} \\
           \nu_{\epsilon_{t_0,sell}} \\
           \nu_{\epsilon_{t_1,buy}} \\
           \nu_{\epsilon_{t_1,sell}} \\
           \vdots \\
           \nu_{\epsilon_{t-1,buy}} \\
           \nu_{\epsilon_{t-1,sell}}
    \end{bmatrix}
    \quad \quad A_{\nu} =
    \begin{bmatrix}
           1.0 & p_{mid,t_0} & S_{ref,t_0} & \epsilon_{t_0, buy} \\
           1.0 & p_{mid,t_0} & S_{ref,t_0} & \epsilon_{t_0, sell} \\
           1.0 & p_{mid,t_1} & S_{ref,t_1} & \epsilon_{t_1, buy} \\
           1.0 & p_{mid,t_1} & S_{ref,t_1} & \epsilon_{t_1, sell} \\
           & & \vdots & \\
           1.0 & p_{mid,t-1} & S_{ref,t-1} & \epsilon_{t-1, buy} \\
           1.0 & p_{mid,t-1} & S_{ref,t-1} & \epsilon_{t-1, sell} \\
    \end{bmatrix}
\end{align*}
similarly for $s_{\epsilon}$,
%\begin{align*} 
%    y = s_{\epsilon} \quad \quad A_{s} = \left[1.0,p_{mid,t},S_{ref,t},\epsilon,\epsilon^2,\epsilon^3\right]
%\end{align*}
\begin{align*}
    y_s = 
    \begin{bmatrix}
           s_{\epsilon_{t_0,buy}} \\
           s_{\epsilon_{t_0,sell}} \\
           s_{\epsilon_{t_1,buy}} \\
           s_{\epsilon_{t_1,sell}} \\
           \vdots \\
           s_{\epsilon_{t-1,buy}} \\
           s_{\epsilon_{t-1,sell}}
    \end{bmatrix}
    \quad \quad A_{s} =
    \begin{bmatrix}
           1.0 & p_{mid,t_0} & S_{ref,t_0} & \epsilon_{t_0, buy} & \epsilon_{t_0, buy}^2 & \epsilon_{t_0, buy}^3\\
           1.0 & p_{mid,t_0} & S_{ref,t_0} & \epsilon_{t_0, sell} & \epsilon_{t_0, sell}^2 & \epsilon_{t_0, sell}^3 \\
           1.0 & p_{mid,t_1} & S_{ref,t_1} & \epsilon_{t_1, buy} & \epsilon_{t_1, buy}^2 & \epsilon_{t_1, buy}^3 \\
           1.0 & p_{mid,t_1} & S_{ref,t_1} & \epsilon_{t_1, sell} & \epsilon_{t_1, sell}^2 & \epsilon_{t_1, sell}^3 \\
           & & & \vdots & & \\
           1.0 & p_{mid,t-1} & S_{ref,t-1} & \epsilon_{t-1, buy} & \epsilon_{t-1, buy}^2 & \epsilon_{t-1, buy}^3 \\
           1.0 & p_{mid,t-1} & S_{ref,t-1} & \epsilon_{t-1, sell} & \epsilon_{t-1, sell}^2 & \epsilon_{t-1, sell}^3 \\
    \end{bmatrix}
\end{align*}

where $m$ is the number of observations and $t$ denotes the time that the order was placed. Notably, The estimator for $s_{\epsilon}$ contains two additional features to model the highly curvilinear relationship between $\epsilon$ and the observed normalized spread profit-and-loss. At inference time $t$, the expected value for net order flow for any $\epsilon_t$ can be computed using the updated estimator $\hat{x}_{\nu,t}$:
\begin{align*} 
    \mathbb{E}(\nu_{\epsilon}) \approx \left[1.0,p_{mid,t},S_{ref,t},\epsilon_t\right]\hat{x}_{\nu,t}
\end{align*}
and similarly for $\mathbb{E}(s_{\epsilon})$. As the model is deployed over time, rows are continually added to the observation matrix $A$, which makes it challenging and cumbersome to continually compute a new estimate $\hat{x}$. To avoid having to compute an increasingly large matrix inversion, the intelligent agent improved its estimates online via the recursive least-squares (RLS) algorithm. %Succinctly, the procedure is as follows:

Similar to the $\mathbb{E}(\nu_{\epsilon})$ and $\mathbb{E}(s_{\epsilon})$ estimates, the optimization procedures also called for variance estimates $var(\nu_{\epsilon})$ and $var(s_{\epsilon})$. These estimates were computed in an online fashion following the procedure given by Knuth \cite{knuth}. Finally, the intelligent agent also kept an estimate of the market volatility given by $\hat{\sigma} = \sigma \sqrt{\Delta p_{mid}}$.

\paragraph{Pricing policy}
The pricing policy follows two steps: at every trade invocation, the intelligent agent chooses price tweaks ($\epsilon_{buy}, \epsilon_{sell}$) to meet its market share target $\eta_{ms}$ and reduce its inventory by skewing its quoted prices.

\textit{Step 1: Meet target market share.} This step ensured that the intelligent agent adapted its policy if it was getting little or no trade flow. Here, the intelligent agent solved for $\epsilon_{*} = \epsilon_{buy} = \epsilon_{sell}$ so that it captured the desired percentage of overall trading volume $\eta_{ms} =$ 25\% within tolerance $\delta_{tol} =$ 5\% \cite{Ganesh2019}. To accomplish this, the intelligent agent used a time-dependent measurement $V_{market, t}$, which was defined as the observed total trading volume between now and the last trade invocation $V_{market, t} = V_{total, t} - V_{total, t-1}$. The initial step of the policy solved for $\epsilon_{*}$ and was formulated as:
\begin{equation*}
    \begin{aligned}
        & \max_{\epsilon}\quad\min_{a} && \left|cost_1(\epsilon) - cost_1(a)\right|\\
        & \textrm{s.t.} && \left|cost_1(\epsilon) - cost_1(a)\right| \le \delta_{tol} \\
        &&& cost_1(\cdot ) = \left| \eta_{ms} - \frac{\mathbb{E(\nu_{\epsilon})}}{V_{market}} \right|\\
    \end{aligned}
\end{equation*}
where $a$ denotes the estimate for $\epsilon$ that achieves the minimal distance from $\eta_{ms}$. The variable $\epsilon$ converges to $\epsilon_{*}$ and maximizes profit whilst remaining close to $\eta_{ms}$.\\
%\paragraph{Step 2: Skew prices to reduce inventory.}
\textit{Step 2: Skew prices to reduce inventory risk.} This step reduced inventory risk by skewing the price of one order side (buy or sell) to attract an order flow that offset the intelligent agent's current inventory $z$. If the inventory $z$ was negative, it decreased $\epsilon_{buy}$ to attract offsetting buy orders and vice versa. Essentially, this step posed a trade-off between spread profit and inventory risk. The final step of the adaptive pricing policy solved for $\epsilon_{skew}$ and was formulated as:
\begin{equation*}
    \begin{aligned}
        & \min_{\epsilon} && -S_{ref}\mathbb{E}(s_{\epsilon)}+\gamma\sqrt{S_{ref}^2 var(s_{\epsilon})+\sigma^2 \mathbb{E}((z+\nu_{\epsilon})^2)}\\
    \end{aligned}
\end{equation*}
where $\gamma = 2$ denotes the risk aversion parameter for the model \cite{Ganesh2019}. After step 2 was completed, the $\epsilon$ value for the side that was being skewed was set to $\epsilon_{skew}$, the other was set to $\epsilon_{*}$, and the limit order was sent over the network. Next, the intelligent agent conducted the hedging policy.
\paragraph{Hedging policy}
Similar to step 2 of the pricing policy, the hedging policy aimed to reduce risk by leveraging opposing trades---only here, instead of setting prices in the hope of attracting an offsetting trade, the intelligent agent directly executed the offsetting trade themselves. The trade-off was between the cost of hedging (by crossing the spread) and inventory risk. The fraction of current inventory to hedge $x$ was formulated as:
\begin{equation*}
    \begin{aligned}
        & \min_{x} && \left| xz \right|S_{ref}+\gamma\sqrt{\sigma^2 \mathbb{E}((z(1-x)+\nu_{\epsilon})^2)}\\
        & \textrm{s.t.} && z(1 - x) = Z \\
        &&& 0 \le x \le 1\\
        &&& -3000 \le Z \le 3000\\
    \end{aligned}
\end{equation*}
where the new inventory $Z$ respects the bounds estimated by empirical high-frequency trading studies \cite{KIRILENKO2017}. The intelligent agent requires more computational resources to operate than the other agent types and relies on a growing design matrix $A$, which inhibits the scalability of the model. In our simulation trials, we only operated a single intelligent agent and ran it on a different machine. Further, we limited this agent to operating on a single asset and excluded it from our multi-asset simulation trials. We leave optimization efforts to the intelligent agent (e.g., exponential forgetting factors, sliding window techniques for growing data matrix, etc. \cite{Vahidi2005, Zhang2000}) and the design of more intelligent agent types (e.g., deep learning agents) to future work.

\section{Results and Discussion}
\label{sec:others}
% In this section, we evaluate key metrics of the baseline financial ABM. First, we describe the calibration of model parameters and then discuss the impact of intelligent behavior on the simulation. Next, we present our ABM's ability to reproduce several well-known stylized facts of financial markets. Finally, we demonstrate the extent to which our ABM can scale in terms of both the population size of agents and the number of available assets for trading.

\subsection{Calibration and stable generation of synthetic financial time series data}

% impact of AMM
\begin{figure}[h!]
     \centering
     \begin{subfigure}{0.49\textwidth}
         \centering
         \includegraphics[width=\textwidth]{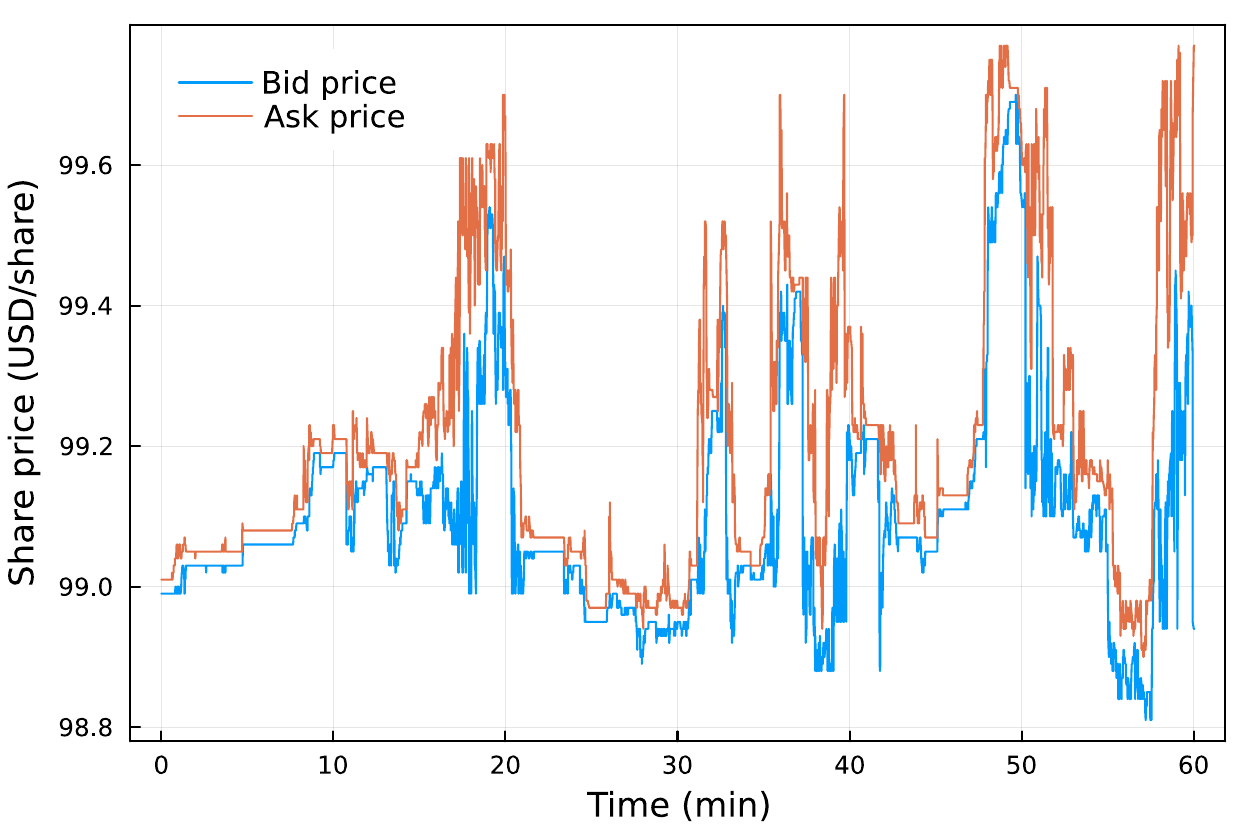}
         %\caption{2 market making agents}
         \caption{}
         \label{fig:ba_2MM}
     \end{subfigure}
     \hfill
     \begin{subfigure}{0.49\textwidth}
         \centering
         \includegraphics[width=\textwidth]{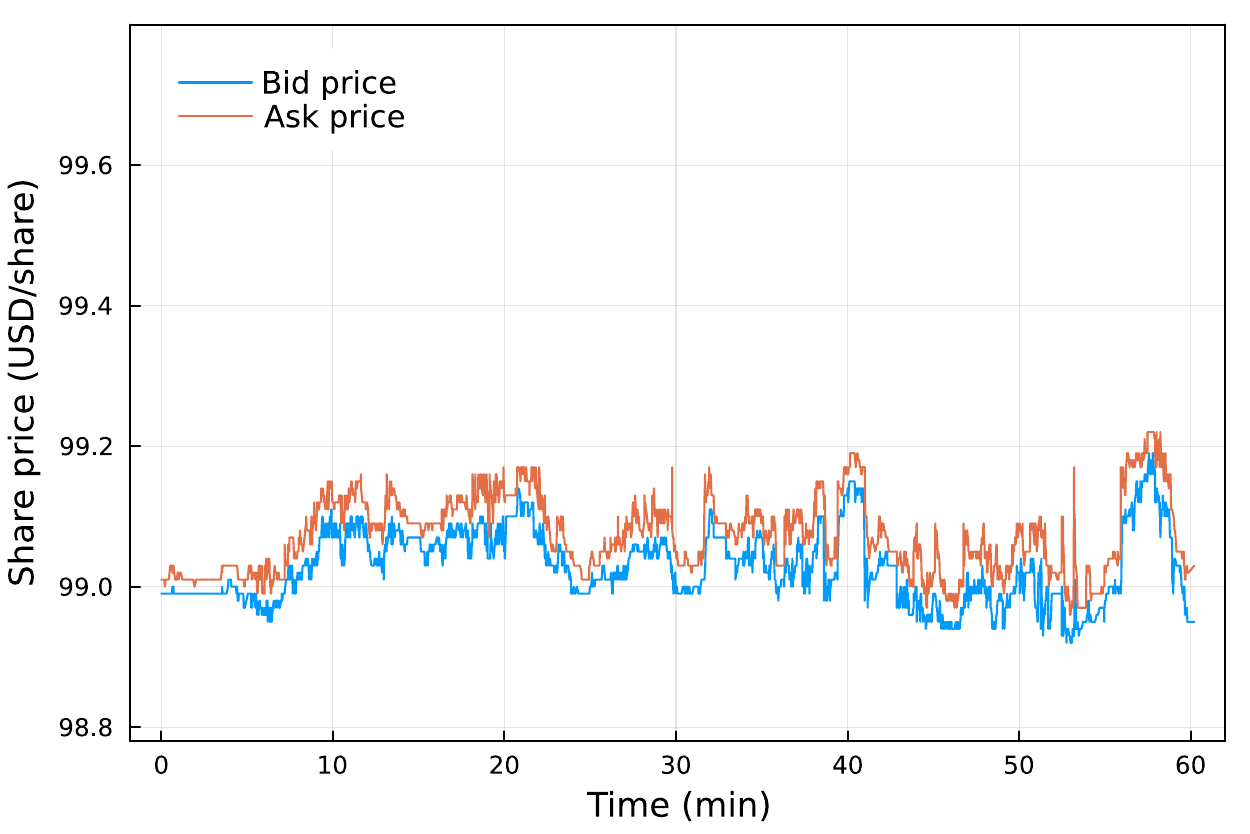}
         %\caption{1 market making agent, 1 intelligent agent}
         \caption{}
         \label{fig:ba_1AMM}
     \end{subfigure}
     \hfill
     \begin{subfigure}{0.49\textwidth}
         \centering
         \includegraphics[width=\textwidth]{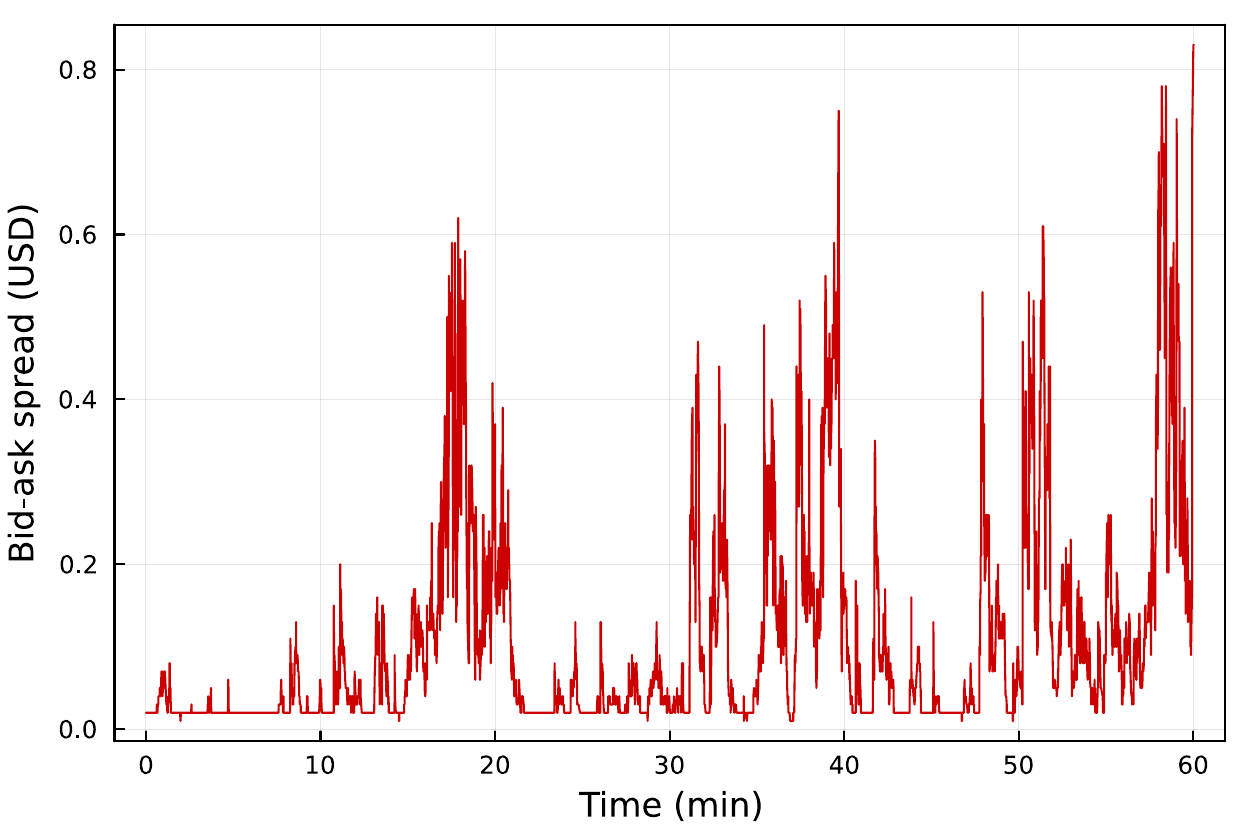}
         %\caption{2 market making agents}
         \caption{}
         \label{fig:spread_2MM}
     \end{subfigure}
     \hfill
     \begin{subfigure}{0.49\textwidth}
         \centering
         \includegraphics[width=\textwidth]{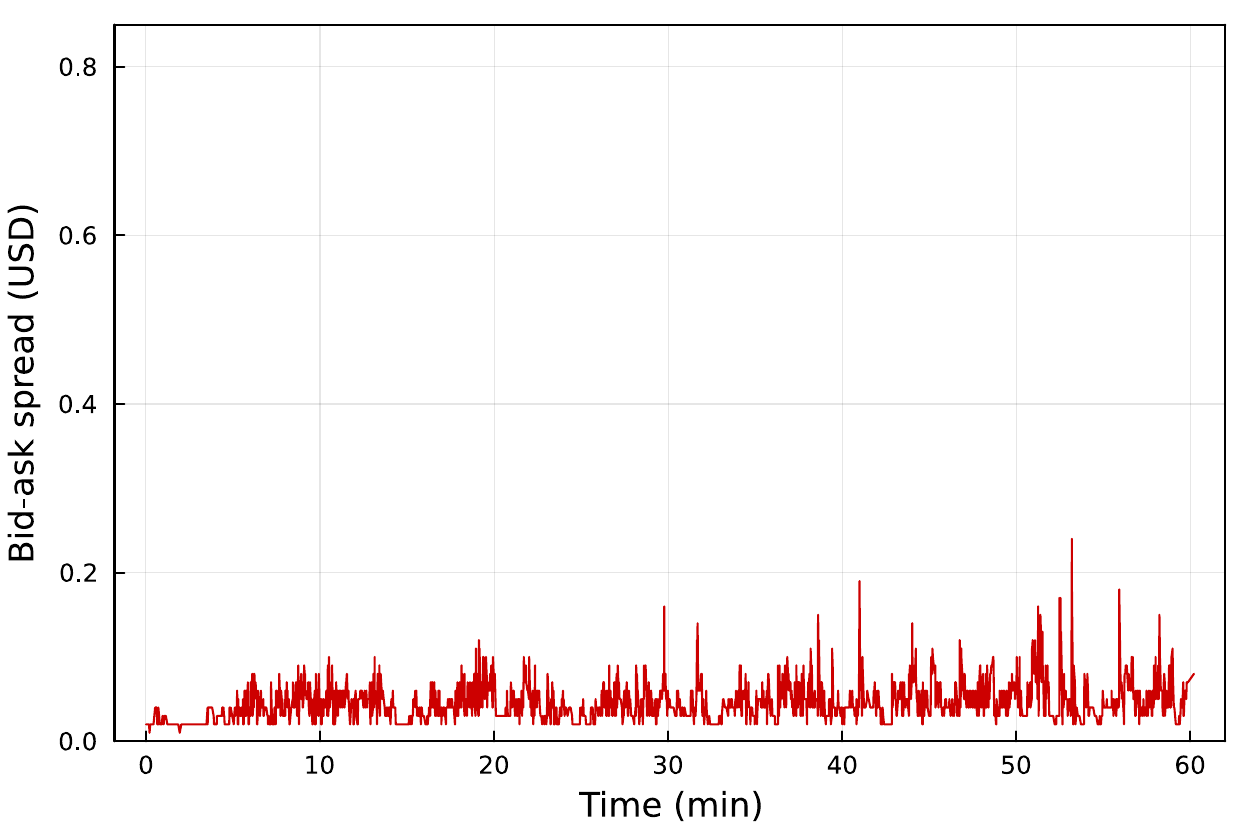}
         %\caption{1 market making agent, 1 intelligent agent}
         \caption{}
         \label{fig:spread_1AMM}
     \end{subfigure}
        \caption{The intelligent agent type effectively stabilized the market environment. (a) Two market-making agents induced relatively volatile bid (blue) and ask (orange) prices, characterized by large upward and downward price movements. (b) One market making agent and one intelligent agent induced stable bid (blue) and ask (orange) prices. (c) Two market-making agents induced large bid-ask spreads (difference between the bid and ask prices) and an overall illiquid market environment characterized by frequent jumps in the cost of trading (crossing the spread). (d) One market-making agent and one intelligent agent induced a stable bid-ask spread and resilient market liquidity.}
        \label{fig:AMM_impact}
\end{figure}

% In agreement with our goal to produce a baseline financial ABM with minimal assumptions, the ABM was designed with as few parameters as possible. 

% The ABM parameters used in this study were estimated using various methods,  some of which depended upon the specific simulation case. 

We considered three simulation cases: small-scale univariate time series generation, medium-scale multivariate time series generation, and large-scale multivariate time series generation (Table~\ref{tab:params}). The liquidity taker and provider agents were defined by four parameters: the number of agents, the initial cash range, the initial shares range, and the agent trading frequency. Each agent's initial cash and share allotment was determined through a draw from a uniform distribution bounded by the range parameters. The market-making agents were defined by four parameters: the number of agents, the price deviation bounds (constant values $\epsilon_{min} = -0.5$ and $\epsilon_{max} = 1.0$ \cite{Ganesh2019}), the unit trade size (constant value of 100, which is a typical lot size \cite{Chakraborti2011}), and the trading frequency. The intelligent agents were defined by six parameters: the market share target ($\eta_{ms} =$ 25\%), risk aversion ($\gamma = 2$ ), tolerance ($\delta_{tol} =$ 5\%), inventory limit (3000), unit trade size (100), and trading frequency. The market-making and intelligent agents were instantiated with zero cash and zero shares (but were allowed to borrow cash and shares).

% 3 cases (parameters influenced by empirical studies X...)
% small number w/ single asset (parameters scaled down)
% medium number of agents and assets (parameters close)
% large number of everything split across several machines (parameters scaled up)
\begin{table}[b]
    \footnotesize
    \centering
    \caption{Model parameters for each simulation case.}
    \begin{tabular}{llll}
         \toprule
         Parameter & Small univariate & Medium multivariate & Large multivariate\\
         \midrule
         Liquidity Takers (LT) & 70 & 650 & 8000 \\
         Liquidity Providers (LP) & 70 & 5850 & 100000 \\
         Market Makers (MM) & 1 & 200 & 500 \\
         Intelligent Agents (IA) & 1 & 0 & 0\\
         Initial cash range & \$5,000 - \$15,000 & \$25,000 - \$75,000 & \$150,000 - \$450,000 \\
         Initial shares range (per asset) & 50 - 150 & 50 - 150 & 50 - 150 \\
         Trading frequencies (seconds) & LT=5, LP=10, MM=2, IA=2 & LT=720, LP=720, MM=2 & LT=7200, LP=7200, MM=2 \\
         Number of unique assets & 1 & 5 & 30\\
         \bottomrule
    \end{tabular}
    \label{tab:params}
\end{table}

We based the simulation's initialization and calibration procedures on previous empirical market studies \cite{Paddrik2012, KIRILENKO2017}. We adjusted these parameters according to our case study's objectives and available computational resources. We found that poorly calibrated parameters led to adverse effects, such as unrealistic liquidity and price movement, which were more noticeable when simulating larger agent populations. Specifically, the simulation performance was sensitive to liquidity takers and providers' initial cash and shares allotment range. If the total initial cash and market value of all assets were roughly equal, the resulting aggregate time series was unbiased. However, if the total initial value of shares was higher than the total initial value of cash, simulation trials were biased toward share prices falling and vice versa. This was a notable feature of our ABM, which suggests that the conservation of total agent cash and shares constraints led to meaningful supply and demand phenomena. To ensure fairness, we used a distribution with a high probability of equal values of total cash and shares in each simulation trial. We set the initial mid-price and random initial order book arrangements for each case study. The parameters for each case study are listed in Table~\ref{tab:params}.

The addition of the intelligent agent effectively stabilized the small-scale univariate simulation environment (Fig. \ref{fig:AMM_impact}). The intelligent agent consistently outperformed other market marking agents and demonstrated a profit in each simulation run. Interestingly, the intelligent agent accomplished this despite the high degree of noise in the environment and the relative simplicity of its linear model. However, upon inspection, we found that the intelligent agent's model produced a poor estimation fit, suggesting that more sophisticated intelligent agents could dominate in this market environment and that more work is needed to produce more realistic odds of profitability in our simulated market. 

\subsection{Validation of synthetic financial time series data}
\begin{figure}[h!]
     %\footnotesize
     \centering
     \begin{subfigure}{0.3\textwidth}
         \centering
         \includegraphics[width=\textwidth]{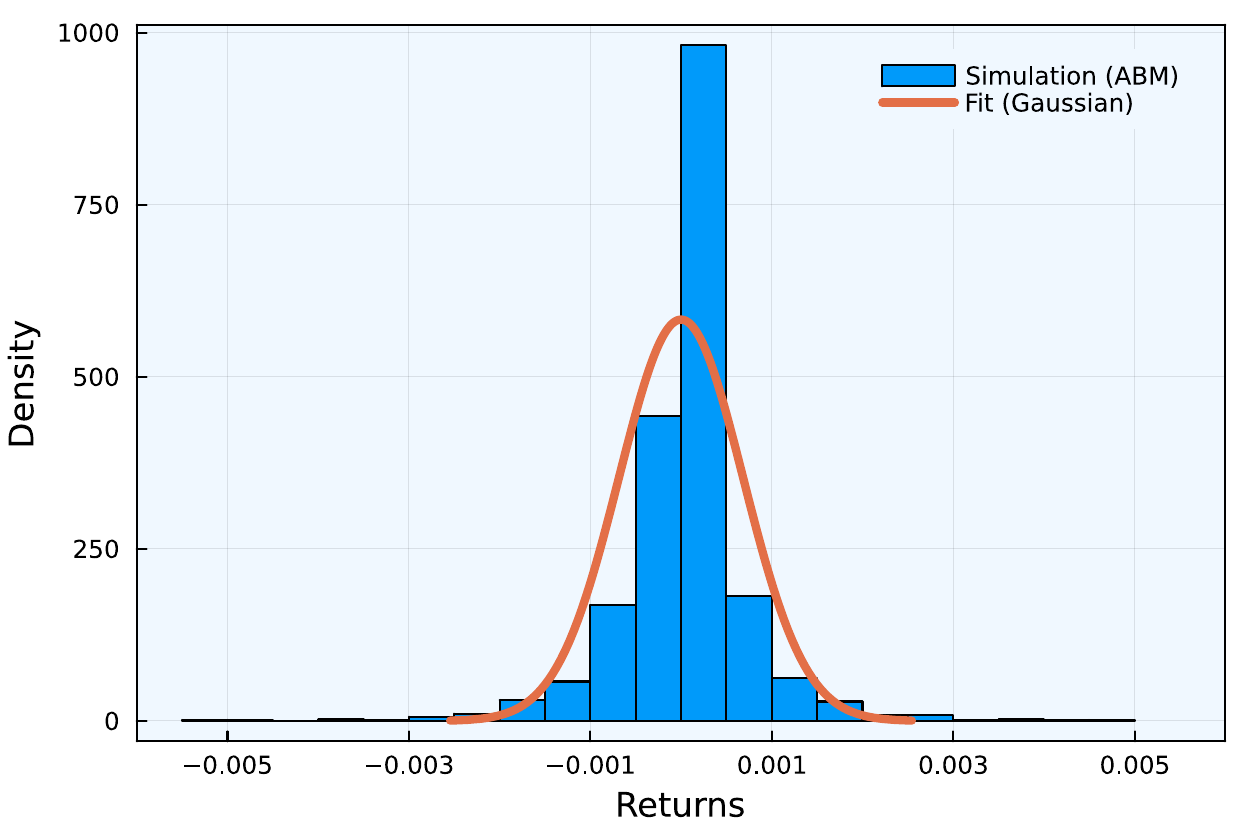}
         %\caption{Heavy tailed distribution}
         \caption{}
         \label{fig:SF1}
     \end{subfigure}
     \hfill
     \begin{subfigure}{0.3\textwidth}
         \centering
         \includegraphics[width=\textwidth]{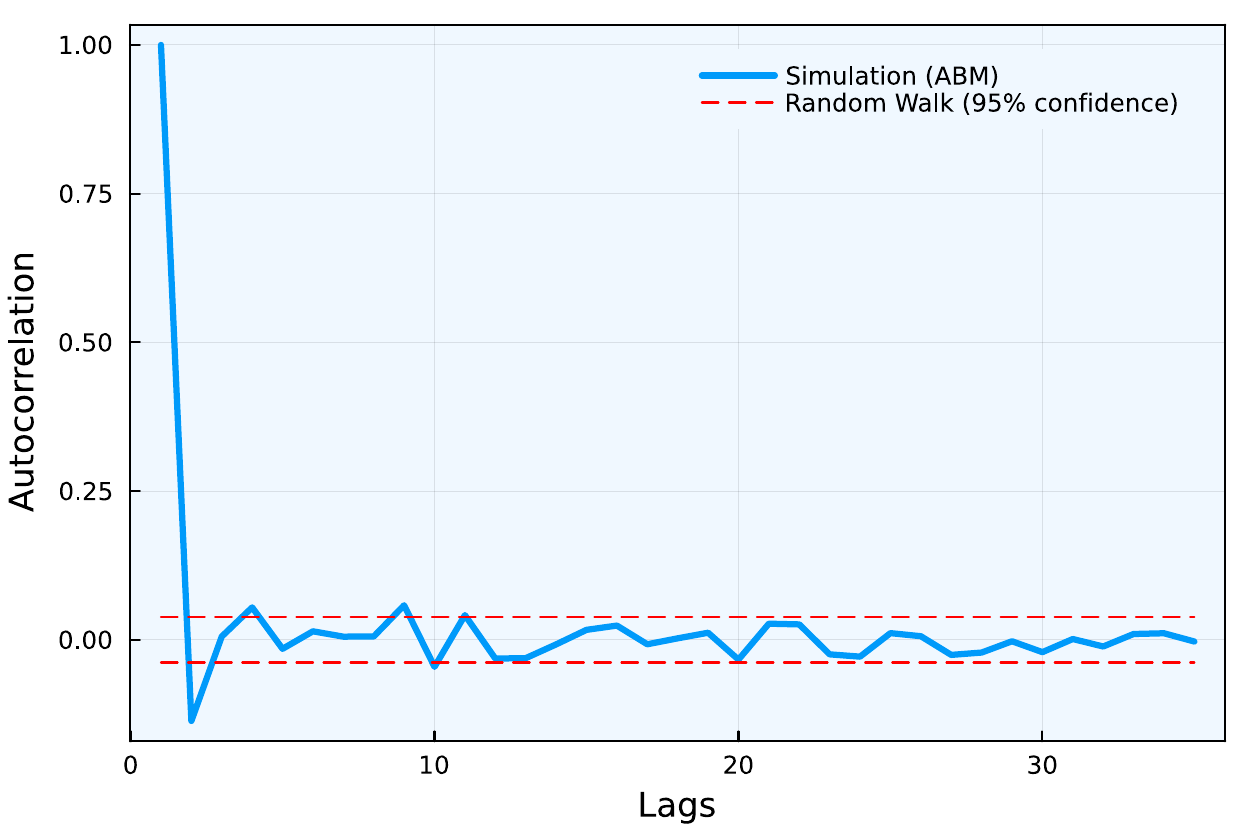}
         %\caption{Linear unpredictability}
         \caption{}
         \label{fig:SF2}
     \end{subfigure}
     \hfill
     \begin{subfigure}{0.3\textwidth}
         \centering
         \includegraphics[width=\textwidth]{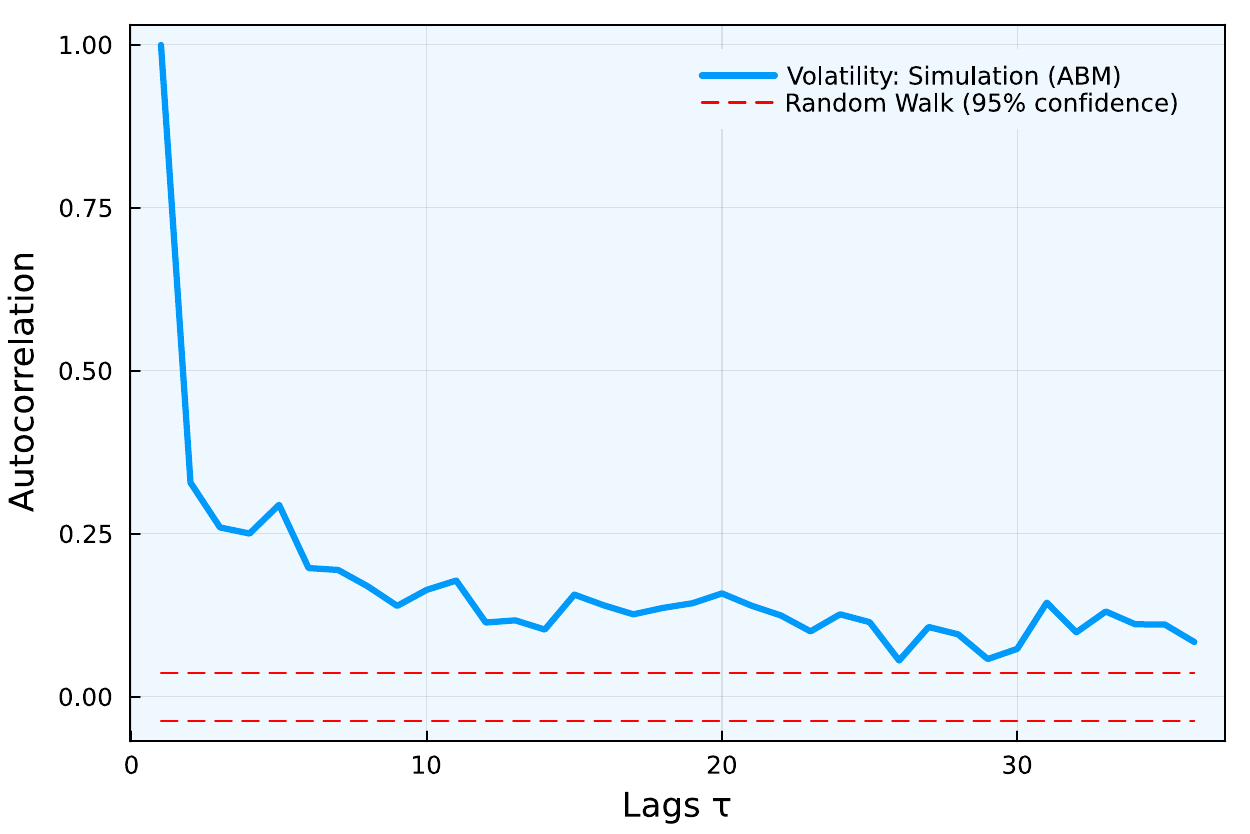}
         %\caption{Volatility clustering}
         \caption{}
         \label{fig:SF3}
     \end{subfigure}
     \hfill
     \begin{subfigure}{0.3\textwidth}
         \centering
         \includegraphics[width=\textwidth]{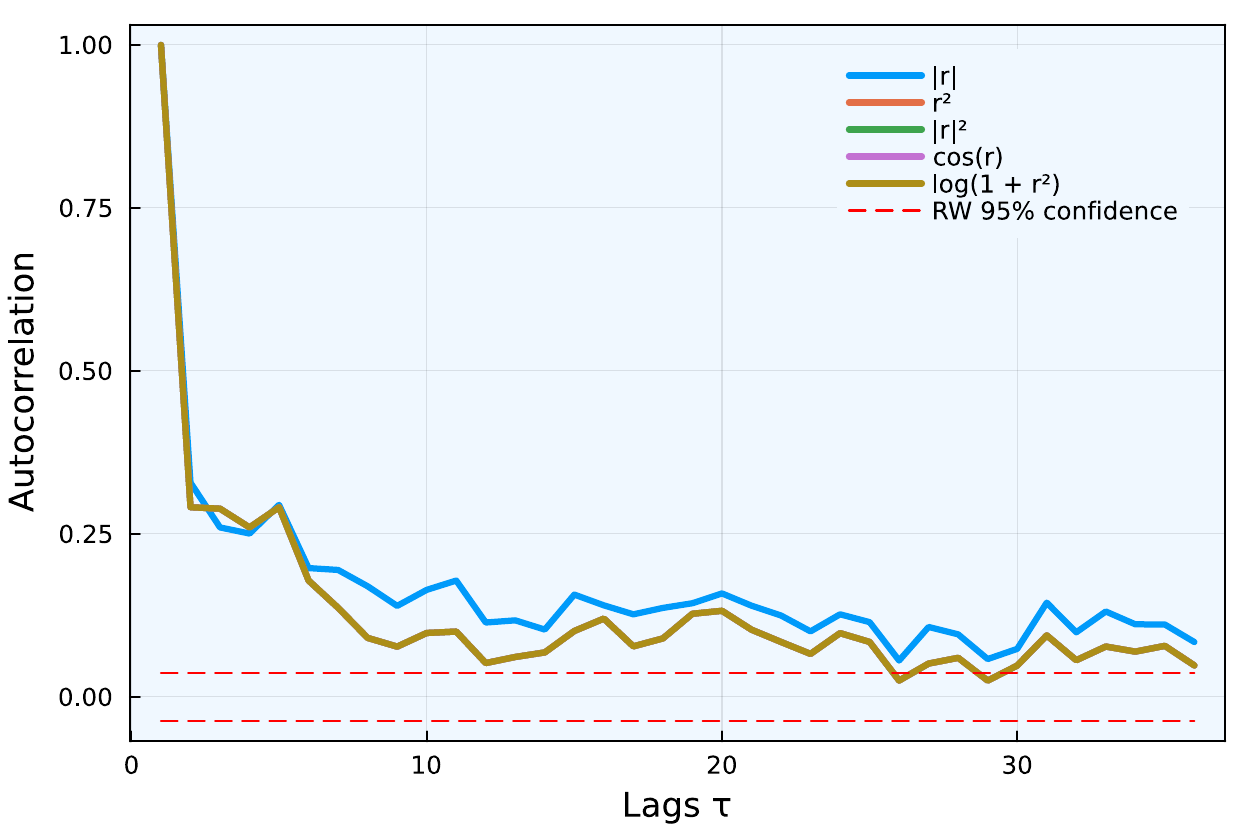}
         %\caption{Nonlinear dependence}
         \caption{}
         \label{fig:SF4}
     \end{subfigure}
     \hfill
     \begin{subfigure}{0.3\textwidth}
         \centering
         \includegraphics[width=\textwidth]{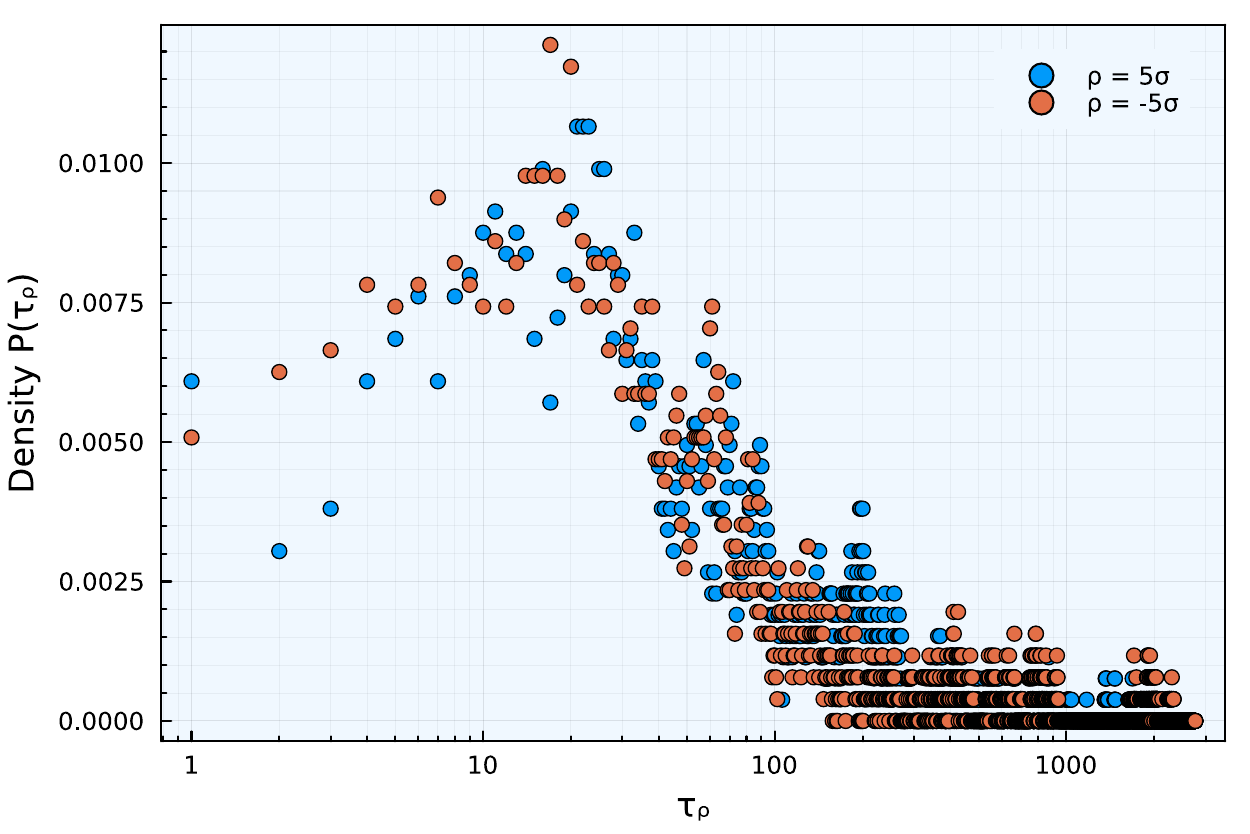}
         %\caption{Gain/loss asymmetry}
         \caption{}
         \label{fig:SF5}
     \end{subfigure}
     \hfill
     \begin{subfigure}{0.3\textwidth}
         \centering
         \includegraphics[width=\textwidth]{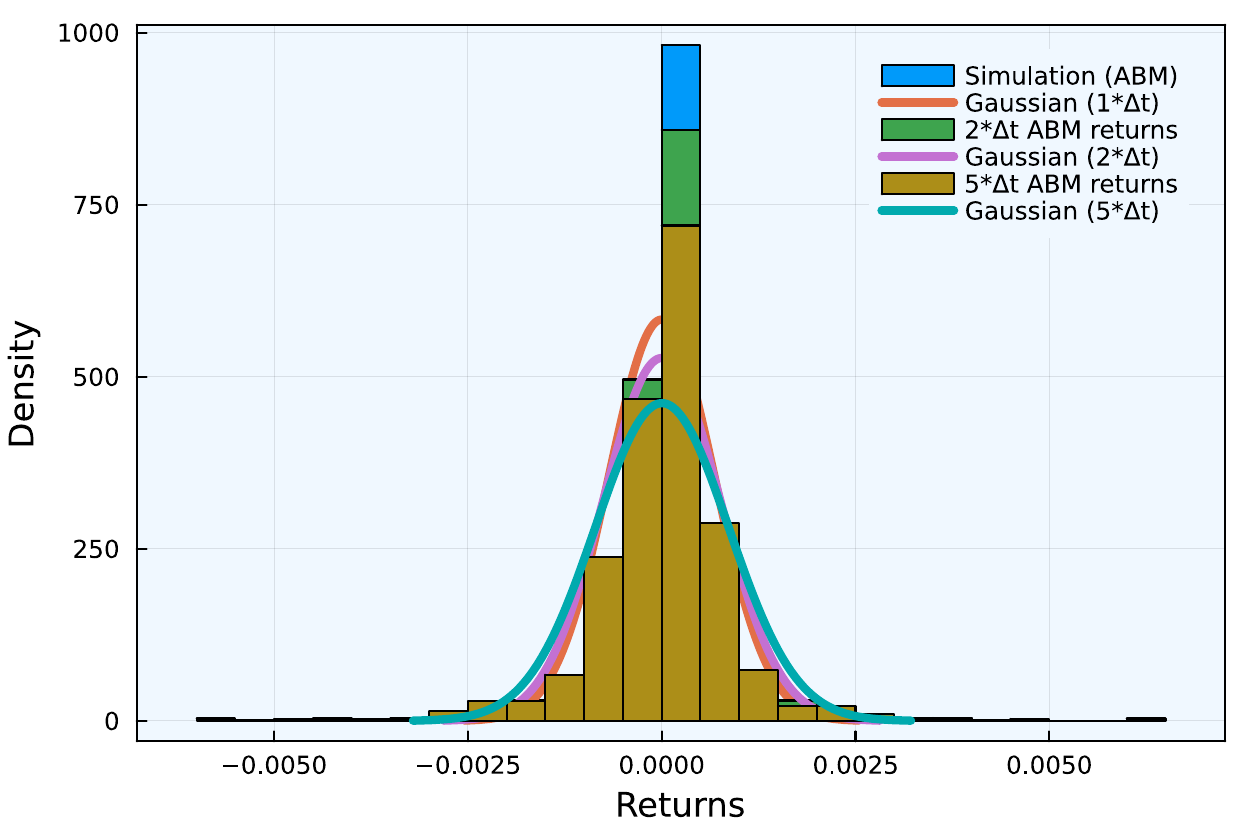}
         %\caption{Aggregational gaussianity}
         \caption{}
         \label{fig:SF6}
     \end{subfigure}
     \hfill
% Subfigure of 10 price series---I want to include, but it's taking up too much room
%     \begin{subfigure}{0.3\textwidth}
%         \centering
%         \includegraphics[width=\textwidth]{Figures/Fig 3/all_trials_prices_1AMM.pdf}
%         \caption{Synthetic price series}
%         \label{fig:10_trials}
%     \end{subfigure}
        \caption{Stylized facts for the time series generated by the baseline financial ABM. (a) Heavy-tailed returns distribution is demonstrated by the high and narrow peak at zero return and the slow decay at the margins of the distribution. (b) Linear unpredictability is demonstrated by the lack of autocorrelation of price returns except at very short time horizons. (c) Volatility clustering is demonstrated by significant positive autocorrelation of price volatility. (d) Nonlinear dependence is demonstrated by significant positive autocorrelation of nonlinear functions of returns (i.e., $|r|$, $r^2$, $|r^2|$, $cos(r)$, and $log(1+r^2)$). (e) Gain/loss asymmetry demonstrated by the distribution of first passage times $\tau_{\rho}$ when an asset crosses a fixed return level $\rho$ (where $\rho$ = $\pm$ 5$\times$ standard deviation of the price series). (f) Aggregational gaussianity is demonstrated by the returns distribution's increasing resemblance to a normal distribution as the time scale over which returns are calculated is increased (i.e., we compare returns calculated from (i) every price tick to (ii) every other price tick and (iii) every fifth price tick). For panels (b)-(d), the autocorrelation series vs. a 95\% confidence interval for a random walk is shown.}
        \label{fig:SF_ABM}
\end{figure}

The baseline financial ABM captured several empirical properties of financial time series (Fig. \ref{fig:SF_ABM}).
To gauge the efficacy of our baseline ABM and validate our approach, we tested our synthetic time series for the presence of several common univariate stylized facts.  We evaluated ten simulation trials of the small-scale univariate case, each sample consisting of an hour's worth of simulated data, and our findings were consistent across all trials. 

% the simulation was unable to capture...
These results indicated that the structure of markets (i.e., double auction mechanism, uncertain order fulfillment, etc.), rather than fully rational agent behavior, was enough to account for several statistical regularities of financial markets. These findings are consistent with other zero-intelligence studies in the literature \cite{FarmerZI2003, Vyetrenko2020}. There are, however, several statistical properties that we were unable to account for using this baseline ABM. Stylized facts that relate to volume (e.g., long memory of volume, volume-volatility correlation) were not captured by our model. This is unsurprising, considering that our baseline agents were not conditioned on volume and were not reactive (i.e., they did not trade/activate based on a conditional signal but rather according to a constant trading frequency). Similarly, the "leverage effect" stylized fact was not captured, and we found this reasonable considering that our zero-intelligence agents possessed no notion of historical performance or personal utility. Further, our baseline agent behaviors were unequipped to replicate stylized facts such as intraday volume patterns (open and close times are not factored into agent behavior) and cross-asset correlations (company sectors are not included in the observation space). The ABM will require extensions to accommodate these and other statistical properties that were unable to be replicated in our baseline implementation. We leave this to future work.

The addition of the intelligent agent had minor impacts on the stylized facts exhibited by the baseline ABM. The only difference was that without the intelligent agent, the generated time series were more volatile, and the presence of stylized facts related to volatility was more pronounced and deviated from what was observed empirically. We suspect that the reason no additional stylized facts were observed with the inclusion of the intelligent agent was likely because its decision-making was conditioned on the same price-centric feature variables as all other agents rather than new ones such as volume and historical trends (e.g., simple moving average and other rolling statistics).

\subsection{Large-scale generation of synthetic financial time series data}
\begin{figure}[h!]
     \centering
     \begin{subfigure}{0.49\textwidth}
         \centering
         \includegraphics[width=\textwidth]{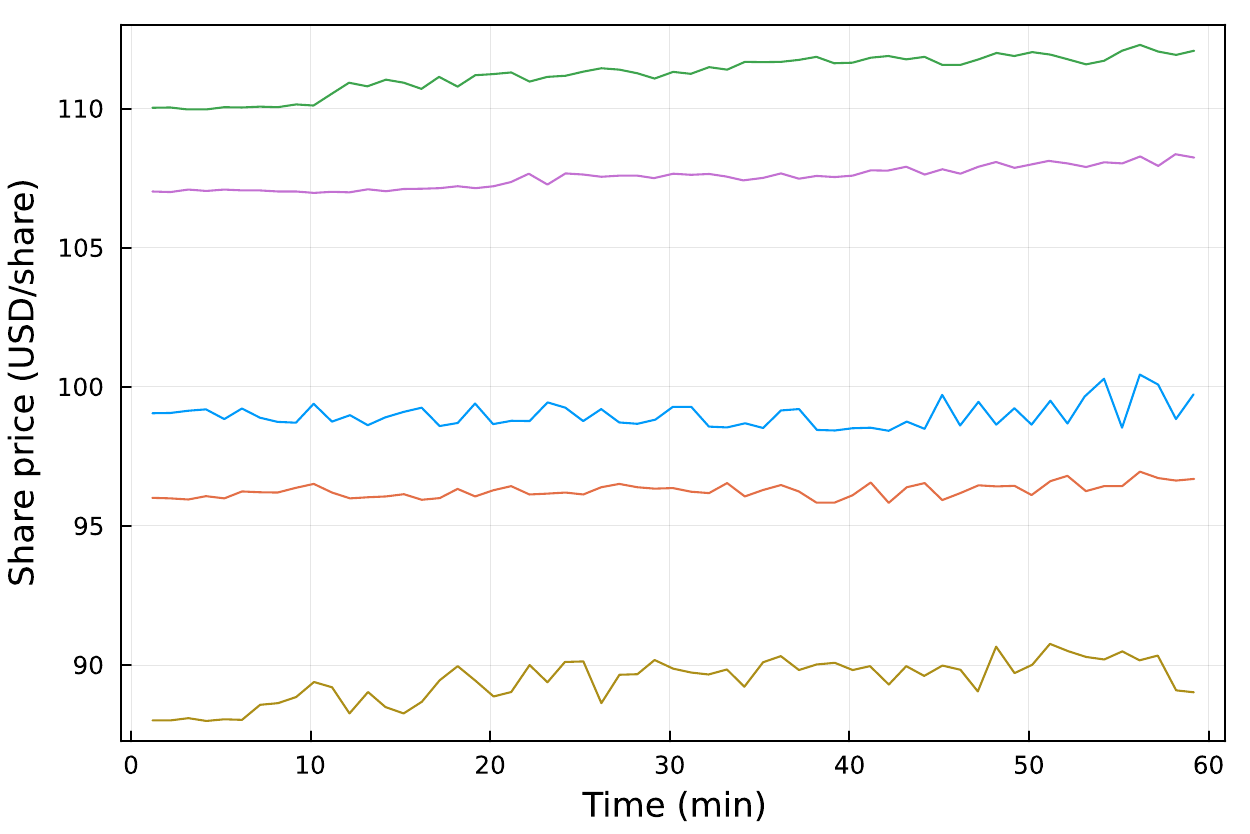}
         \caption{}
         \label{fig:5tick_price}
     \end{subfigure}
     \hfill
     \begin{subfigure}{0.49\textwidth}
         \centering
         \includegraphics[width=\textwidth]{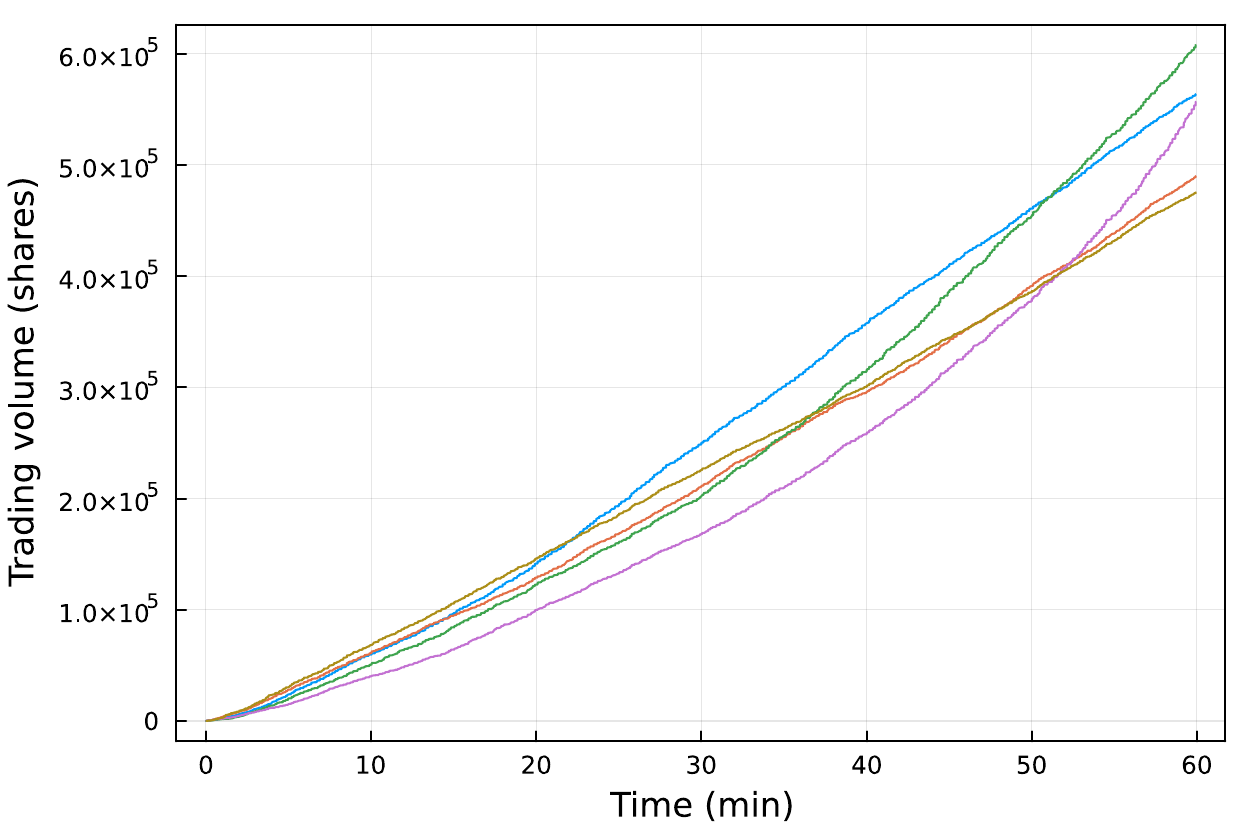}
         \caption{}
         %\label{fig:5tick_volume}
     \end{subfigure}
     \begin{subfigure}{0.49\textwidth}
         \centering
         \includegraphics[width=\textwidth]{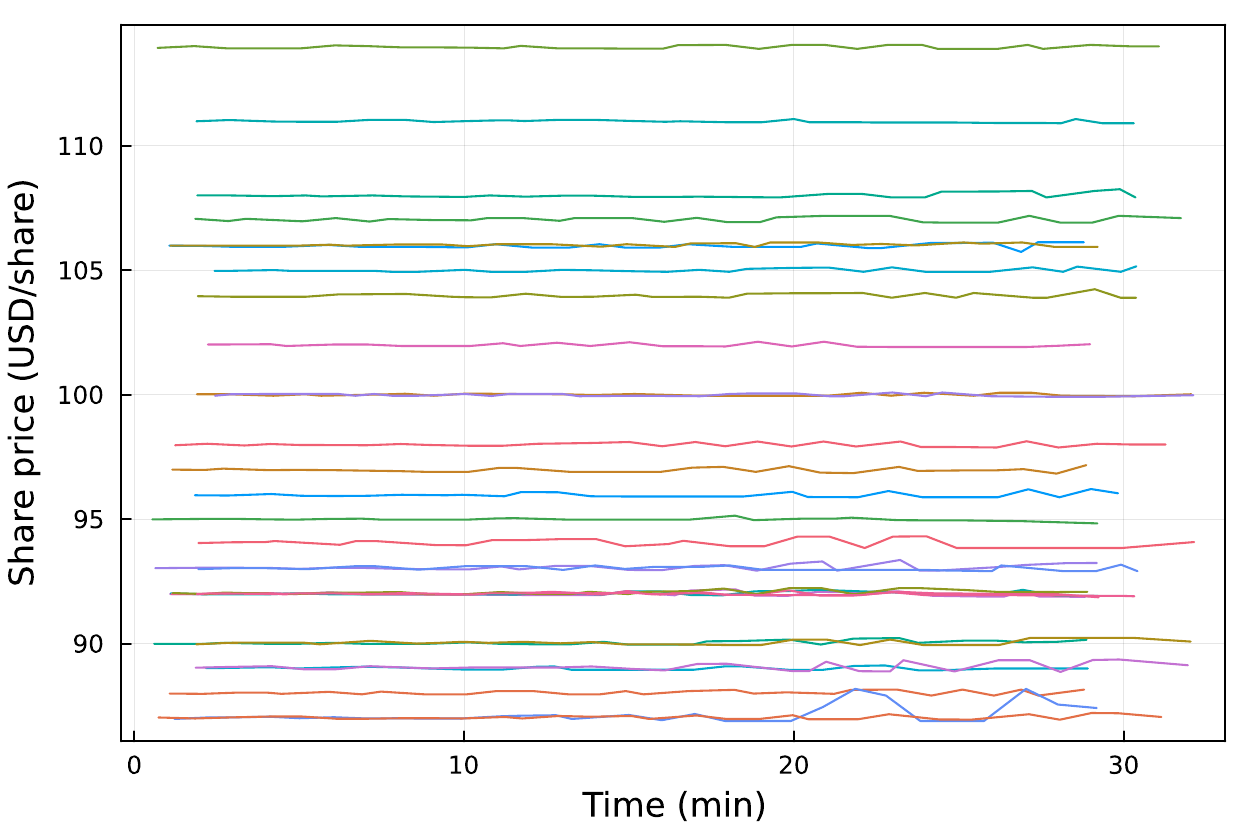}
         \caption{}
         \label{fig:30tick_price}
     \end{subfigure}
     \hfill
     \begin{subfigure}{0.49\textwidth}
         \centering
         \includegraphics[width=\textwidth]{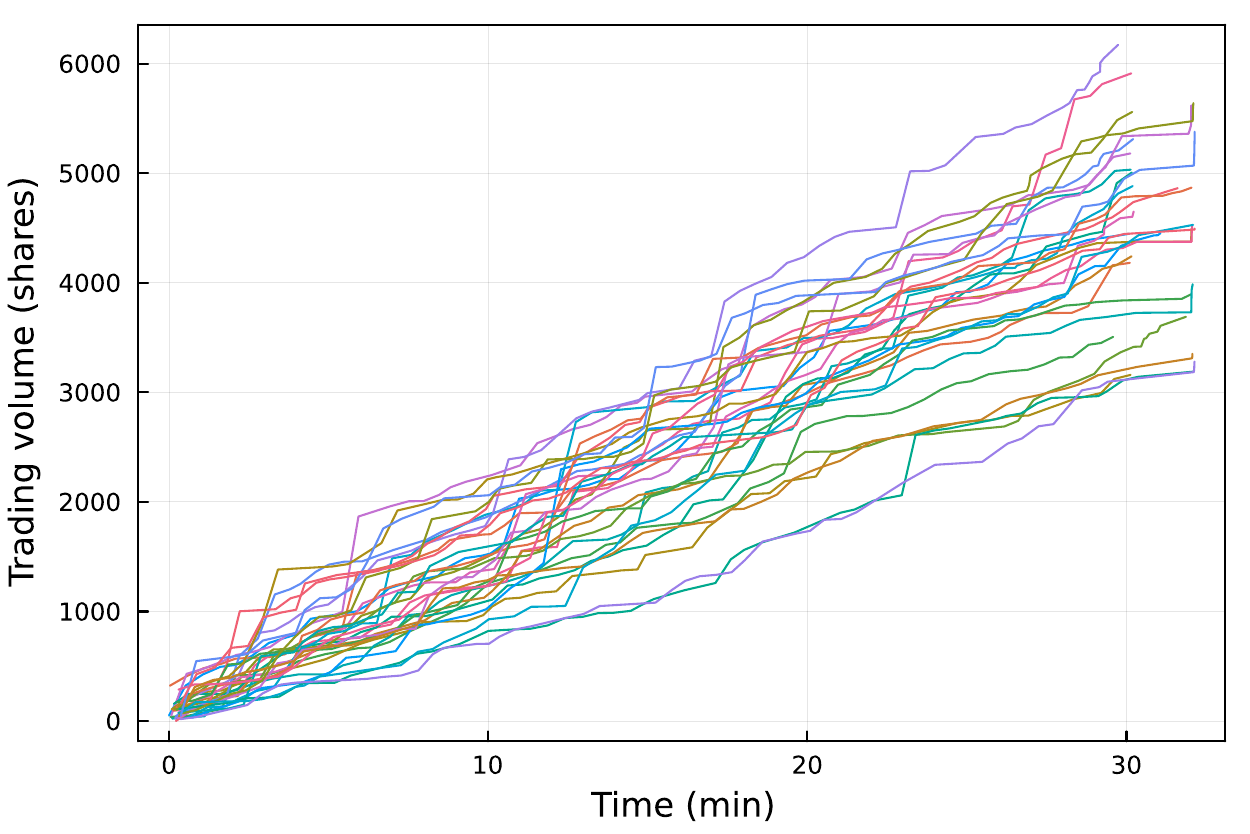}
         \caption{}
         \label{fig:30tick_volume}
     \end{subfigure}
        \caption{Price and volume series for the multivariate simulation trials. (a) Synthetic price series for each of the five assets of the medium-scale multivariate simulation case. (b) Synthetic market volume series or cumulative shares traded, for each of the five assets of the medium-scale multivariate simulation case. (c) Synthetic price series for each of the 30 assets of the large-scale multivariate simulation case. (d) Synthetic market volume series or cumulative shares traded for each of the 30 assets of the large-scale multivariate simulation case.}
        \label{fig:large_scale_trials}
\end{figure}

Our large-scale simulation trials consisted of multiple unique assets and several thousand agents operating across multiple machines (Fig. ~\ref{fig:large_scale_trials}). For the large-scale trials, we split the central server and agents across three machines that were each equipped with an Intel Core i7-6700 CPU and Ubuntu 22.04. Our machine allotment for each simulation component was based on minimizing computational bottlenecks: the central system (server hosting order books and observable variables) was allotted its own machine, the market-making agents (high-frequency market intermediaries) operated on a single machine, and the liquidity providers and takers (low-frequency traders) were both operated on the single remaining machine.

% include panels for small-scale univariate for comparison?

The initial share prices of the assets were drawn from a uniform distribution bounded by \$85.00 and \$115.00. Our results demonstrated how the baseline ABM could support multiple simultaneous assets by only using basic agent types and modest computational resources. With greater resources, it would be straightforward to scale the simulation even further--opening up possibilities to explore multivariate stylized facts and investigate more complicated agent behaviors. For instance, a large-scale ABM would allow researchers to investigate portfolio management agent behaviors and trading differences between small-tick and large-tick stocks.

Although the baseline ABM was capable of supporting large agent populations and many available assets, the scale of trading activity (i.e., executed orders or transactions) was hindered by the use of a single central system for both order processing and retrieving observable information. For instance, after 30 minutes, the medium-scale simulation trial produced approximately 45,000 trades across all assets, and the large-scale simulation trial produced only 2,000. We suspect that this is due to the large message load induced on the single gateway to the central server and the look-up times associated with the large SQLite database. To simulate realistic large-scale trading activity, components of the baseline ABM will have to be further optimized and partitioned (e.g., the communication should be optimized, and the brokerage functionality of the central system should be isolated from the exchange/order books segment). We leave this to future work.

\section{Summary and Conclusion}
This study introduced a baseline financial ABM that scaled to several thousand agents, supports multiple simultaneous assets, captured intuitive supply and demand dynamics, and demonstrated several well-known statistical properties of financial markets. Further, our ABM does all of this without the inclusion of traditional agent behaviors (e.g., fundamental, technical, etc.) or the need for historical market data. That being said, the ABM can be readily extended to support additional agent types and will need to do so in order to improve on realism and capture additional stylized facts. In particular, our results emphasize the need to include conditional activation times for agents and agent behaviors that are conditioned on variables such as historical price performance metrics, market sectors, and time of day. We cannot conclude that intelligent agent behaviors are needed for the reproduction of stylized facts, but our results do suggest that intelligent agent behavior may be needed for resilient market liquidity. Moreover, we suspect that the inclusion of several intelligent agent types and competition among them will be necessary to produce more realistic probability odds in the simulation. Additional ways to improve simulation fidelity of the baseline ABM, especially at large scale, include the reorganization of simulation components and the adoption of low-latency and high-throughput communication frameworks.

% discuss cons/limitations of our system here? Or discuss this in beginning of methods section? Conclusion? Cons include speed and nondeterminism
% Future work could include... more order types such as stop-loss, more sophisticated agents, fragmented exchange with different matching engines, AMM does not scale to multiple assets among other issues

% \section*{Acknowledgments}
% This was was supported in part by......

\section*{Data availability statement}
This work was produced using the Brokerage.jl and TradingAgents.jl Julia packages. Model codes are available at: \url{https://github.com/aaron-wheeler/Brokerage.jl.git} and \url{https://github.com/aaron-wheeler/TradingAgents.jl.git}.

\clearpage

%Bibliography
\bibliographystyle{unsrt}  
\bibliography{references}  

\clearpage

% \appendix
\section*{Appendix A. Trading agent algorithms}\label{app:agents}
Here, we provide pseudo-code for the agents in the simulation environment. 
Agent behaviors followed one of four types: liquidity takers (Algorithm \ref{alg:LT}), 
liquidity providers (Algorithm \ref{alg:LP}), market makers (Algorithm \ref{alg:MM}), 
and intelligent market makers (Algorithm \ref{alg:IA}).

\SetKwComment{Comment}{/* }{ */}

\begin{algorithm}
    \caption{Liquidity taker agent algorithm}\label{alg:LT}
    \KwData{Market state ($t, p_{mid}$), \\ \qquad~~ agent parameters ($c_{min}, c_{max}, h_{min}, h_{max}, t_{freq}$), \\ \qquad~~ and simulation parameters ($N, K, T_{close}$)}
    \KwResult{Consistent market order (MO) flow for duration of simulation}
    \BlankLine
    \tcc{Initialize agents}
    % Initialize agents \\
    \For{agent $n \in N$}{
    $c_{n, t_0} \gets U(c_{min}, c_{max})$\;
    \lForEach{asset $k \in K$}{$h_{n, t_0}^k \gets U(h_{min}, h_{max})$}  
    }
    \BlankLine
    \tcc{Execute trades}
    \While{$t < T_{close}$}{
        \For{agent $n \in N$}{
        Sample $u_{n, t}$ $\sim$ U(0, 1) \\
          \If{$u_{n, t}$ < $\frac{1}{t_{freq}}$}{
            Sample risk fraction $\mathcal{X}_{n,t}$ $\sim$ U(0, 1) \\
            $\hat{\mathcal{W}}^{R}_{n,t} \gets \mathcal{X}_{n,t} (c_{n,t} + \sum_{k=1}^{K} h^k_{n,t}\cdot p^k_{mid, t})$\;
            Sample portfolio weights $\hat{\omega}_{n,t+1}^k$ $\sim$ D($K$, 1.0) \\
            \For{asset $k \in K$}{
                $\hat{h}_{n,t+1}^k \gets \left\lfloor \hat{\omega}_{n,t+1}^k \cdot
                \frac{\hat{\mathcal{W}}^R_{n,t}}{p_{mid, t}^k} \right\rfloor$ \;
                $\Delta^k_{n,t} \gets \hat{h}_{n,t+1}^k - h_{n,t}^k$ \;
                {\eIf{$\Delta^k_{n,t} < 0$}{
                    Submit buy side MO with volume $\Delta^k_{n,t}$\;
                    }{
                    Submit sell side MO with volume $\Delta^k_{n,t}$\;
                    }
                }
            }    
          }
        }
    }
\end{algorithm}

\begin{algorithm}
    \caption{Liquidity provider agent algorithm}\label{alg:LP}
    \KwData{Market state ($t, p_{bid}, p_{ask}, p_{h}: \forall h=0,1,\ldots,t$), \\ \qquad~~ agent parameters ($c_{min}, c_{max}, h_{min}, h_{max}, t_{freq}$), \\ \qquad~~ and simulation parameters ($N, K, T_{close}$)}
    \KwResult{Consistent limit order (LO) flow for duration of simulation}
    \BlankLine
    \tcc{Initialize agents}
    % Initialize agents \\
    \For{agent $n \in N$}{
    $c_{n, t_0} \gets U(c_{min}, c_{max})$\;
    \lForEach{asset $k \in K$}{$h_{n, t_0}^k \gets U(h_{min}, h_{max})$}  
    }
    \BlankLine
    \tcc{Execute trades}
    \While{$t < T_{close}$}{
        \lForEach{asset $k \in K$}{$p_{mid, t}^k \gets \frac{p_{bid, t}^k + p_{ask, t}^k}{2}$}
        \For{agent $n \in N$}{
            Sample $u_{n, t}$ $\sim$ U$(0, 1)$ \\
            \If{$u_{n, t}$ < $\frac{1}{t_{freq}}$}{
                \For{asset $k \in K$}{
                $r_{t}^k \gets \sum_{t=1}^{T} ln(p_{t}^k) - ln(p_{t-1}^k)$\;
                $\mu \gets \frac{1}{T} \sum_{t=1}^{T}(r_{t}^k)$\;
                $\sigma_{p_{mid, t}^k} \gets \sqrt{\frac{1}{T} \sum_{t=1}^{T} (r_t - \mu)^2}$\;
                Sample $\mathcal{X}_{n,t}\sim \mathcal{N}\left(0,\sigma_{p_{mid, t}^k} \right)$\;
                $p_{n,t}^{L_k} \gets p_{mid, t}^k (1+\mathcal{X}_{n,t})$\;
                }
                \For{asset $k \in K$}{
                    \If{$h^k_{n,t} > 0$ {\bf and} $p_{mid} < p_{n,t}^{L_k}$}{
                    $\Delta^k_{n,t} \gets U(0,1) \ast h^k_{n,t}$\;
                    Submit sell side LO with volume $\Delta^k_{n,t}$ and limit price $max\left\{p_{ask, t},p_{n,t}^{L_k}\right\}$\;
                    }
                }
                \For{eligible assets $k_e \in K_e$ where $K_e = count(c_{n,t} > p_{mid, t}^k)$}{
                Sample cash allocation for eligible assets $\hat{c}_{n,t}^k$ $\sim$ D$(K_e, 1.0)$\;
                    \For{eligible assets $k_e \in K_e$}{
                    $\Delta^k_{n,t} \gets \left\lfloor \hat{c}_{n,t}^k \cdot
                    \frac{c_{n,t}}{\text{min}\left\{p_{bid, t},p_{n,t}^{L_k}\right\}} \right\rfloor$ \;
                    Submit buy side LO with volume $\Delta^k_{n,t}$ and limit price $min\left\{p_{bid, t},p_{n,t}^{L_k}\right\}$\;
                    }
                }   
          }
        }
    }
\end{algorithm}

\begin{algorithm}
    \caption{Market making agent algorithm}\label{alg:MM}
    \KwData{Market state ($t, p_{mid}, S_{ref}$), \\ \qquad~~ agent parameters ($\epsilon_{min}, \epsilon_{max}, O_{size}, t_{freq}$), \\ \qquad~~ and simulation parameters ($N, K, T_{close}$)}
    \KwResult{Consistent liquidity across assets for duration of simulation}
    \BlankLine
    \tcc{Initialize agents}
    \For{agent $n \in N$}{
    $c_{n, t_0} \gets 0$\;
    \lForEach{asset $k \in K$}{$z_{n, t_0}^k \gets 0$}  
    }
    \BlankLine
    \tcc{Execute trades}
    \While{$t < T_{close}$}{
        \For{agent $n \in N$}{
        Sample $u_{n, t}$ $\sim$ U$(0, 1)$ \\
        \If{$u_{n, t}$ < $\frac{1}{t_{freq}}$}{
            \For{asset $k \in K$}{
                $\epsilon_{buy}^k \gets U(\epsilon_{min}, \epsilon_{max})$\;
                $\epsilon_{sell}^k \gets U(\epsilon_{min}, \epsilon_{max})$\;
                $p^{bid_k}_t \gets p_{mid, t}^k + S_{ref,t}^k (1 + \epsilon_{buy}^k)$\;
                $p^{ask_k}_t \gets p_{mid, t}^k - S_{ref,t}^k (1 + \epsilon_{sell}^k)$\;
                Submit buy and sell side LOs with volume $O_{size}$ and limit price $p^{bid_k}_t$ and $p^{ask_k}_t$, respectively\;
                \Begin{
                Sample hedge fraction $\mathcal{X}_{n,t}$ $\sim$ $U(0, 1)$\;
                $O_{hedge} \gets \mathcal{X}_{n,t} \ast z_{n, t}^k$\;
                \eIf{$z_{n, t}^k > 0$}{
                Submit sell side MO with volume $O_{hedge}$\;
                }{
                Submit buy side MO with volume $O_{hedge}$\;
                }
                }
                \If{Order is not filled}{
                Cancel all unfilled buy and sell LOs\;
                }
                Update cash $c_{n, t}$ and inventory $z_{n, t}^k$\;
            }    
          }
        }
    }
\end{algorithm}

\begin{algorithm}
    \caption{Intelligent agent algorithm}\label{alg:IA}
    \KwData{Market state ($t, p_{mid}, S_{ref}, V_{total}$), \\ \qquad~~ agent parameters ($\eta_{ms},\gamma,\delta_{tol},Z,O_{size},t_{freq}$), \\ \qquad~~ and simulation parameters ($T_{close}$)}
    \KwResult{Strategic trading activity for single asset for duration of simulation}
    \SetKwFunction{PricingPolicyTarget}{PricingPolicyTarget}
    \SetKwFunction{PricingPolicySkew}{PricingPolicySkew}
    \SetKwFunction{HedgePolicy}{HedgePolicy}
    \SetKwFunction{RLS}{RLS}
    \SetKwFunction{Wait}{Wait}
    \BlankLine
    \tcc{Initialize agent}
    $c_{t_0} \gets 0$\;
    $z_{t_0}^k \gets 0$\;
    \BlankLine
    \tcc{Execute trades for single asset}
    \While{$t < T_{close}$}{
        $\hat{\sigma} \gets \sigma\sqrt{p_{mid,t} - p_{mid,t-1}}$\;
        $V_{M} \gets V_{total,t} - V_{total,t-1}$\;
        $\epsilon_{\ast} \gets$ \PricingPolicyTarget{$\eta_{ms},\delta_{tol},V_{M},\mathbb{E}(\nu_{\epsilon})$}\;
        $\epsilon_{skew} \gets$ \PricingPolicySkew{$S_{ref},\mathbb{E}(s_{\epsilon}),\gamma,var(s_{\epsilon}),\sigma,var(\nu_{\epsilon})$}\;
        \eIf{$z > 0$}{
        $p^{bid}_t \gets p_{mid, t} + S_{ref,t} (1 + \epsilon_{\ast})$\;
        $p^{ask}_t \gets p_{mid, t} - S_{ref,t} (1 + \epsilon_{skew})$\;
        }{
        $p^{bid}_t \gets p_{mid, t} + S_{ref,t} (1 + \epsilon_{skew})$\;
        $p^{ask}_t \gets p_{mid, t} - S_{ref,t} (1 + \epsilon_{\ast})$\;
        }
        Submit buy and sell side LOs with volume $O_{size}$ and limit price $p^{bid}_t$ and $p^{ask}_t$, respectively\;
        Solve for hedge fraction $\mathcal{X}_{t} \gets$ \HedgePolicy{$z_t,S_{ref},\gamma,\sigma,var(\nu_{\epsilon}),Z$}\;
        $O_{hedge} \gets \mathcal{X}_{t} \ast z_{t}^k$\;
        \eIf{$z_t > 0$}{
        Submit sell side MO with volume $O_{hedge}$\;
        }{
        Submit buy side MO with volume $O_{hedge}$\;
        }
        \While{$V_{total,t} == V_{total,t-1}$}{
        \Wait{$t_{freq}$}\;
        }
        \If{Order is not filled}{
        Cancel all unfilled buy and sell LOs\;
        }
        Update cash $c_{t}$ and inventory $z_{t}$\;
        Update observation matrices $A_{\nu}$ and $A_s$\;
        Update measurement vectors $\nu_t$ and $s_t$\;
        $\hat{x}_{\nu,t}, P_{\nu,t}, K_{\nu,t} \gets$ \RLS{$\hat{x}_{\nu,t-1}, P_{\nu,t-1}, K_{\nu,t-1}, A_{\nu}, \nu_t$}\;
        $\hat{x}_{s,t}, P_{s,t}, K_{s,t} \gets$ \RLS{$\hat{x}_{s,t-1}, P_{s,t-1}, K_{s,t-1}, A_s, s_t$}\;
    }
\end{algorithm}

\end{document}